\documentclass[11pt]{article}
\usepackage{amsmath,amssymb,color,graphics,epsfig,cite}
\usepackage{hyperref}
\hypersetup{hidelinks}
\usepackage{ulem}
\usepackage{graphicx} 
\usepackage{tikz}
\usepackage{array}
\usepackage{tabularx}
\usepackage{booktabs}
\usepackage[section]{placeins}
\usepackage{amsfonts}

\newcommand{\be}{\begin{equation}}
\newcommand{\ee}{\end{equation}}
\newcommand{\bea}{\setlength\arraycolsep{2pt} \begin{eqnarray}}
\newcommand{\eea}{\end{eqnarray}}
\newcommand{\nn}{\nonumber}

\def\ft#1#2{{\textstyle{\frac{\scriptstyle #1}{\scriptstyle #2} } }}
\def\fft#1#2{{\frac{#1}{#2}}}

\def\0{{\sst{(0)}}}
\def\1{{\sst{(1)}}}
\def\2{{\sst{(2)}}}
\def\3{{\sst{(3)}}}
\def\4{{\sst{(4)}}}
\def\5{{\sst{(5)}}}
\def\6{{\sst{(6)}}}
\def\7{{\sst{(7)}}}
\def\8{{\sst{(8)}}}
\def\sst#1{{\scriptscriptstyle #1}}

\begin{document}

 
\begin{center}
		{\Large {\bf Echoes and light rings of regular black holes and their singular counterparts}}

		\vspace{40pt}
  
{\large Xiao-Ping Rao\footnote{xiaopingrao@jxnu.edu.cn;}, Xiao Qian Wang and Hyat Huang\footnote{hyat@mail.bnu.edu.cn;} }

\vspace{10pt}

{\it School of Physics, \\
	Jiangxi Normal University, Nanchang~330022, Jiangxi Province, China}

\vspace{40pt}
		
		\underline{ABSTRACT}
	\end{center}

Recent work [Phys. Rev. D 111, 104040] showed that the Bardeen and Hayward solutions arise as fine-tuned regular members of broader families that also contain singular black holes. We study massless scalar and auxiliary Maxwell perturbations of these solutions using time-domain evolution and a sixth-order WKB approximation. For part of the singular branch, the time-domain profiles exhibit late-time echoes and the effective potentials develop a double-barrier structure. We also examine null geodesics of the background metric. The same singular branch can have two exterior maxima of the null effective potential separated by a minimum, whereas the type-I regular black holes have a single exterior maximum. The Bardeen and Hayward families display the same qualitative behavior. 

~\\
~\\
~\\
~\\
~\\
~\\
~\\

~\\
~\\
~\\
~\\

	\thispagestyle{empty}

	\tableofcontents
\addtocontents{toc}{\protect\setcounter{tocdepth}{2}}
	

 
 \newpage

\section{Introduction}

Regular black holes, namely black holes without curvature singularities, play an important role in gravitational physics. The first concrete model was proposed by Bardeen in 1968 \cite{bardeen} and was later obtained within the framework of nonlinear electrodynamics (NLED) \cite{Ayon-Beato:1998hmi}. Since then, regular black-hole solutions have been constructed in general relativity (GR) coupled to various matter sources \cite{Bronnikov:2000vy,Fan:2016hvf,Li:2023yyw,Li:2024rbw,Bokulic:2022cyk,Bronnikov:2005gm,Bronnikov:2012ch,Bueno:2021krl,Lavrelashvili:1992ia}, in modified theories of gravity \cite{Balart:2014cga,Nojiri:2017kex,Berej:2006cc,Junior:2015fya,Rodrigues:2015ayd,Ghosh:2018bxg,deSousaSilva:2018kkt,Bueno:2024dgm,Koshelev:2024wfk}, and in approaches to quantum gravity \cite{Ashtekar:2023cod,Lewandowski:2022zce,Zhang:2024ney,Bueno:2025gjg}. Regular geometries have also been obtained using the Hadamard partie finie technique \cite{Capozziello:2024ucm}.

Regular black holes have been studied in connection with singularity theorems \cite{Penrose:1964wq,DeFelice:2024seu,An:2021plu,An:2022lvo,Yang:2021civ,Cai:2020wrp}, primordial black holes \cite{Dialektopoulos:2025mfz,Davies:2024ysj,Calza:2024xdh,Calza:2024fzo}, and the AdS/CFT correspondence \cite{Laassiri:2023azi,ElMoumni:2021zbp,Lin:2016swr}. The detection of gravitational waves from binary black-hole mergers has also made their dynamical properties directly relevant to observation \cite{LIGOScientific:2016aoc,LIGOScientific:2016dsl}. Several studies have considered regular black holes as possible sources of gravitational waves \cite{Kumar:2024our,Gong:2023ghh,Cao:2024oud,Yang:2024lmj}. A coalescing binary evolves through inspiral, merger, and ringdown. During ringdown, the remnant is characterized by its quasinormal modes (QNMs), whose frequencies depend on the background spacetime. QNMs have consequently been computed for a wide variety of compact objects \cite{Pedrotti:2024znu,Bianchi:2021xpr,Khoo:2024yeh,Blazquez-Salcedo:2022pwc,Blazquez-Salcedo:2021exm,Blazquez-Salcedo:2018pxo,Koshelev:2024lyu}. Reviews of the subject can be found in Refs.~\cite{Kokkotas:1999bd,Berti:2009kk,Konoplya:2011qq}.

The QNMs of regular black holes have been studied extensively \cite{Fernando:2012yw,Flachi:2012nv,Ulhoa:2013fca,Toshmatov:2015wga,Konoplya:2022hll,Wu:2024ldo,Tang:2024txx,Tang:2025mkk,Huang:2023aet}. Most of these studies compare regular solutions with familiar singular solutions of GR, such as the Schwarzschild and Reissner--Nordstr\"om black holes. It is also useful to compare regular and singular black holes within the same theory. Such a comparison requires a theory whose solution space contains both branches.

Such a setting was recently constructed in GR \cite{Huang:2025uhv}. When NLED is used as the matter source, a regular black hole can occur at a fine-tuned point in a larger solution family that also contains singular black holes. The Bardeen and Hayward solutions provide two examples: the regular member can have two horizons, while a singular member of the same family can have three. Here we investigate the QNMs of these solutions while keeping the underlying theory fixed.

We find late-time echoes for part of the singular branch. In these cases, the perturbation potential has two barriers. The relation between such potentials and circular null orbits \cite{Vagnozzi:2022moj,Cardoso:2016rao,Hashimoto:2023buz,Huang:2021qwe,Ou:2021efv,Pedrotti:2025upg,Tsukamoto:2020iez} motivates us to examine the null geodesics and light-ring structure of the same solutions.

The paper is organized as follows. Section~\ref{2} reviews the Bardeen and Hayward families constructed in Ref.~\cite{Huang:2025uhv}. In Sec.~\ref{3}, we study massless scalar and auxiliary Maxwell perturbations by time-domain evolution and the WKB method. Section~\ref{4} discusses null geodesics and light rings of the background metric. We summarize our results in Sec.~\ref{5}. The main text focuses on the Bardeen family; the corresponding Hayward results are collected in Appendices A and B.

\section{Regular black holes and their singular counterparts}\label{2}

We begin by reviewing the main result of Ref.~\cite{Huang:2025uhv}. Consider a theory with action
\be\label{Lag}
S=\int d^4 x\,\sqrt{-g}\,{\cal L}= \int d^4 x\,\sqrt{-g}\,\big(R+{\cal L}_M\big),
\ee
where $R$ is the Ricci scalar and ${\cal L}_M$ is the matter Lagrangian; we use units in which $16\pi G=1$. We consider the static, spherically symmetric line element
\be\label{ds}
ds^2=-f(r)dt^2+\frac{dr^2}{f(r)}+r^2(d\theta^2+\sin^2\theta\,d\varphi^2).
\ee
Reference~\cite{Huang:2025uhv} showed that, if $f_0(r)$ is a regular black-hole solution of the theory \eqref{Lag}, the same theory also admits the one-parameter family
\be
f(r)=f_0(r)-\ft{2m}{r}.
\ee
For $m\ne0$, the additional term introduces a curvature singularity at $r=0$.
Varying $m$ moves through a family of distinct static solutions of the same theory. It does not describe the time-dependent formation or evolution of an individual black hole.

Two well-known regular black holes arise in NLED. The Lagrangian and metric function of the Bardeen family are\footnote{We define $F=dA$ and $F^2=F_{\mu\nu}F^{\mu\nu}$.}
\be
{\cal L}_B=\ft{1}{\alpha}\ft{(\alpha F^2)^\fft{5}{4}}{(1+\sqrt{\alpha F^2})^\fft{5}{2}}, \qquad f(r)_B= 1-\ft{r^2}{6\alpha (1+r^2)^{\fft{3}{2}}}  .
\ee

The full family admitted by the same theory is
\be\label{fbf}
f(r)_{BF}= 1-\ft{r^2}{6\alpha (1+r^2)^{\fft{3}{2}}}-\ft{2m}{r}.
\ee
For the Hayward family, the corresponding expressions are
\be
{\cal L}_H=\ft{1}{\alpha} \ft{(\alpha F^2)^\fft{3}{2}}{(1+(\alpha F^2)^\fft{3}{4})^2}, \qquad f(r)_H= 1-\ft{r^2}{6\alpha(1+r^3)}  .
\ee
The full Hayward family in the same theory is
\be\label{hbf}
f(r)_{HF}=1-\ft{r^2}{6\alpha(1+r^3)}-\ft{2m}{r}.
\ee
We take the coupling constant $\alpha$ to be positive. The two families have similar properties, so we use the Bardeen family to present the analysis in the main text.

At large $r$, Eq.~\eqref{fbf} has the expansion
\be
f_{BF}|_{r\to \infty}=1-\frac{2m+\frac{1}{6\alpha}}{r}+\frac{1}{4\alpha r^3}-\frac{5}{16\alpha r^5}+{\cal O}(r^{-7}) .
\ee
The ADM mass is therefore
\be
M=m+\ft{1}{12 \alpha}.
\ee
Positivity of the mass requires $m>-\ft{1}{12\alpha}$.

Near the center, Eq.~\eqref{fbf} behaves as
\be
f_{BF}|_{r\to 0}=1-\ft{2m}{r}-\ft{r^2}{6\alpha}+\ft{r^4}{4\alpha}+\dots \: .
\ee
The metric is regular at the center only when $m=0$, or equivalently when $M=\ft{1}{12\alpha}$. The singularity is timelike for $m<0$ and spacelike for $m>0$.

\begin{figure}[!htbp]
\centering
\includegraphics[width=0.45\textwidth]{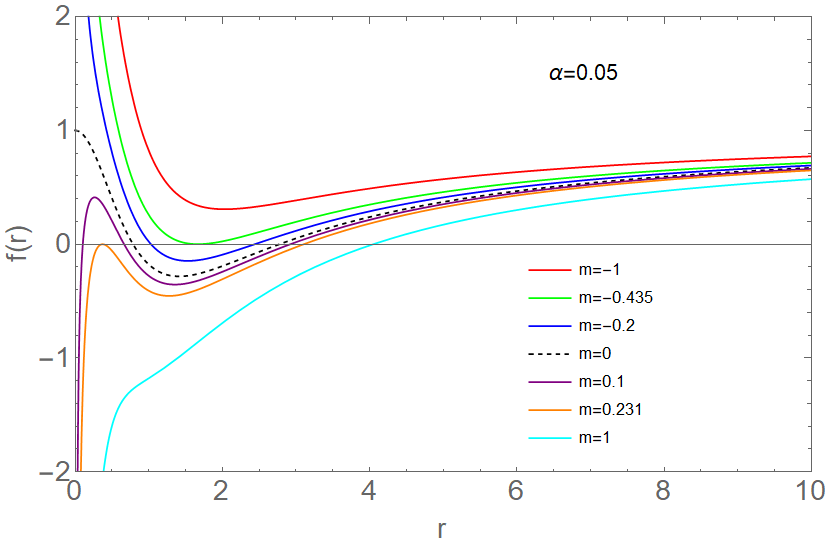} \qquad
\includegraphics[width=0.45\textwidth]{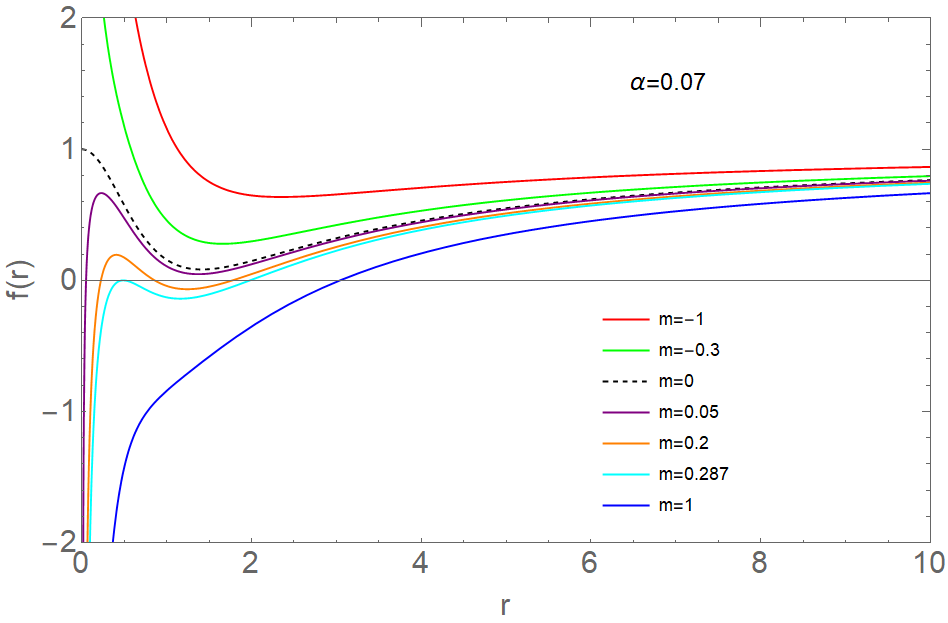}
\caption{ \it Metric functions obtained by varying $m$ away from the regular member at fixed $\alpha=0.05$ (left) and $\alpha=0.07$ (right). The black dashed curves denote a regular two-horizon black hole and a regular horizonless geometry with a de Sitter core, respectively. Static members with $m<0$ have a timelike singularity at $r=0$, whereas those with $m>0$ have a spacelike singularity.}
\label{Bhxiao}
\end{figure}

For $m=0$, the metric describes a black hole when $f_B(r)=0$ has a positive real root. The three cases are
\begin{itemize}
    \item for $\alpha < \ft{1}{9\sqrt{3}}$, the regular state is a two-horizon black hole;
    \item for $\alpha = \ft{1}{9\sqrt{3}}$, the regular state is an extremal black hole with $r_h=\sqrt{2}$;
    \item for $\alpha > \ft{1}{9\sqrt{3}}$, the regular state is a horizonless geometry with a de Sitter core.
\end{itemize}
Representative metric functions are shown in Fig.~\ref{Bhxiao}. In the left panel, the regular member is already a black hole. Its outer horizon grows continuously as $m$ increases. We refer to this family as \textit{type-I black holes}.

In the right panel, the regular member is a horizonless geometry with a de Sitter core. At sufficiently large positive $m$, the corresponding static solution has a new outer horizon and therefore describes a black hole. The outer-horizon radius changes discontinuously along this parameter-space branch. We refer to this family as \textit{type-II black holes}.

\section{Quasinormal modes and echoes}\label{3}

We now compute the QNMs of a massless scalar field and an auxiliary Maxwell field on the fixed black-hole backgrounds. The Maxwell field used here is minimally coupled and should not be confused with a perturbation of the NLED field that supports the background; a complete NLED analysis would in general involve coupled matter and metric perturbations. Previous studies computed the QNMs of the regular Bardeen and Hayward solutions while varying the couplings of the theory \cite{Fernando:2012yw,Flachi:2012nv,Ulhoa:2013fca,Toshmatov:2015wga}. Here we instead hold the couplings fixed and vary the solution parameter $m$, so that all configurations belong to the same theory. We present the Bardeen results in this section and the corresponding Hayward results in Appendix A.

\FloatBarrier
\subsection{The Schr\"odinger-like equation}
\suppressfloats[t]

\begin{figure}[!htbp]
\centering
\includegraphics[width=0.40\textwidth]{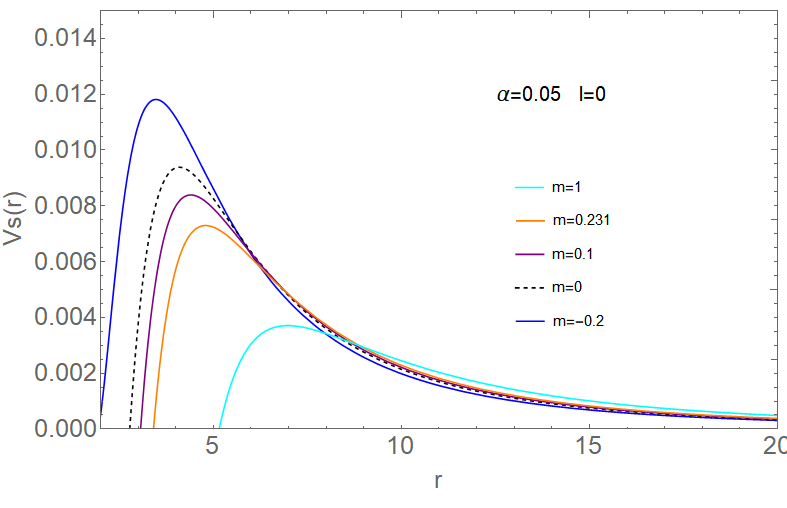} \qquad
\includegraphics[width=0.40\textwidth]{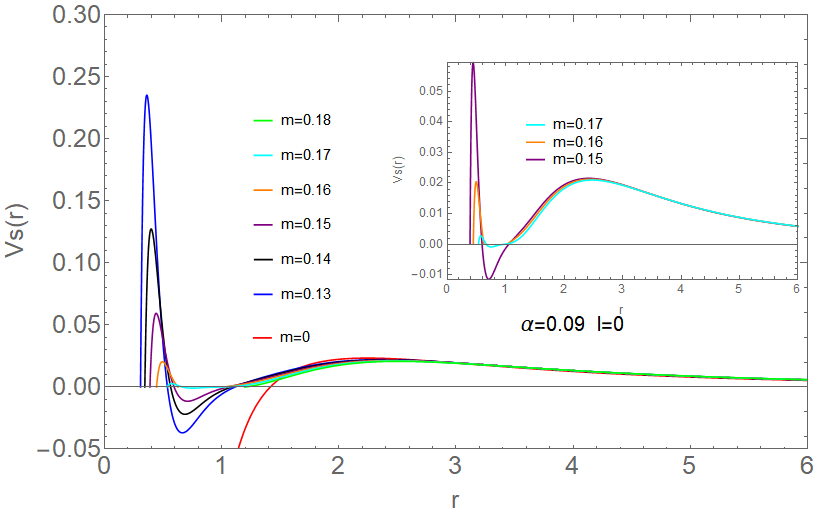}
\includegraphics[width=0.40\textwidth]{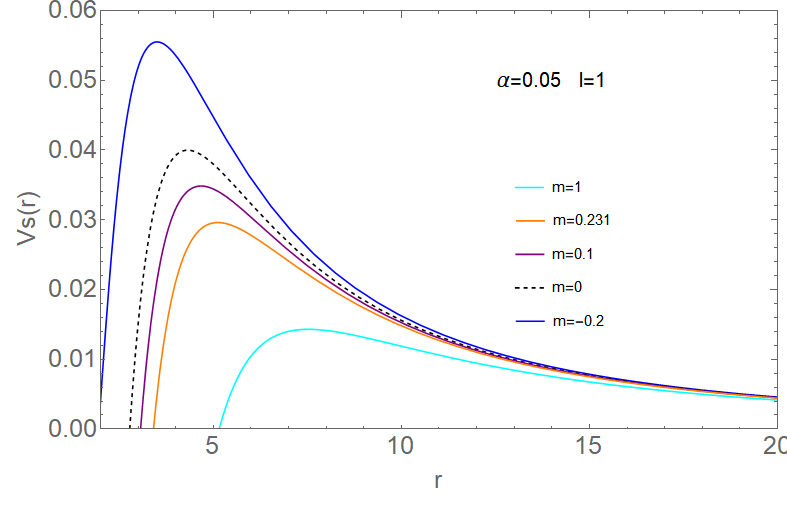} \qquad
\includegraphics[width=0.40\textwidth]{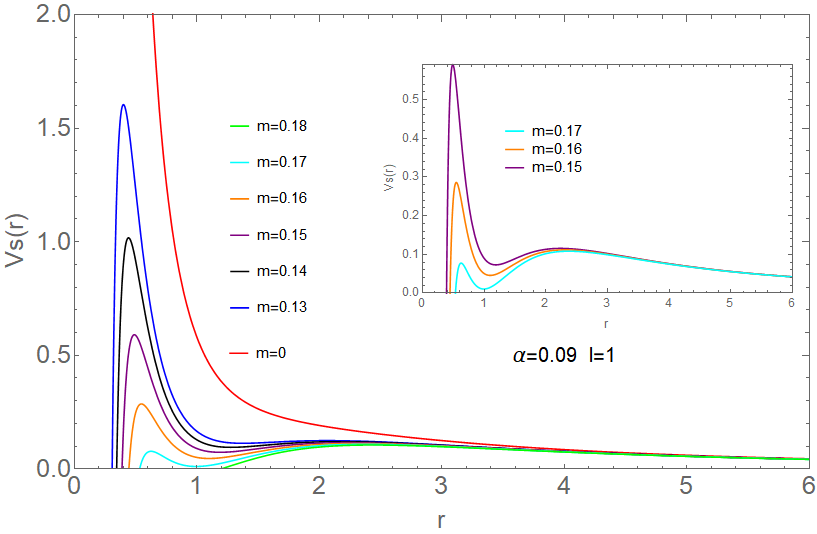}
\includegraphics[width=0.40\textwidth]{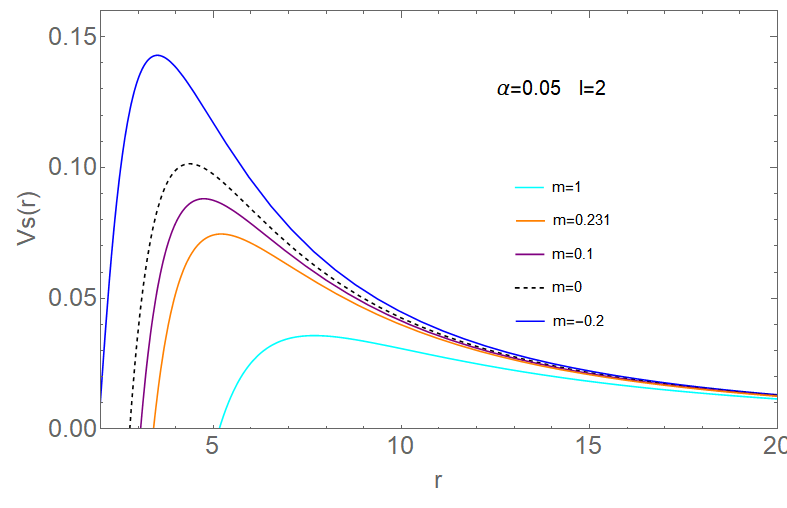} \qquad
\includegraphics[width=0.40\textwidth]{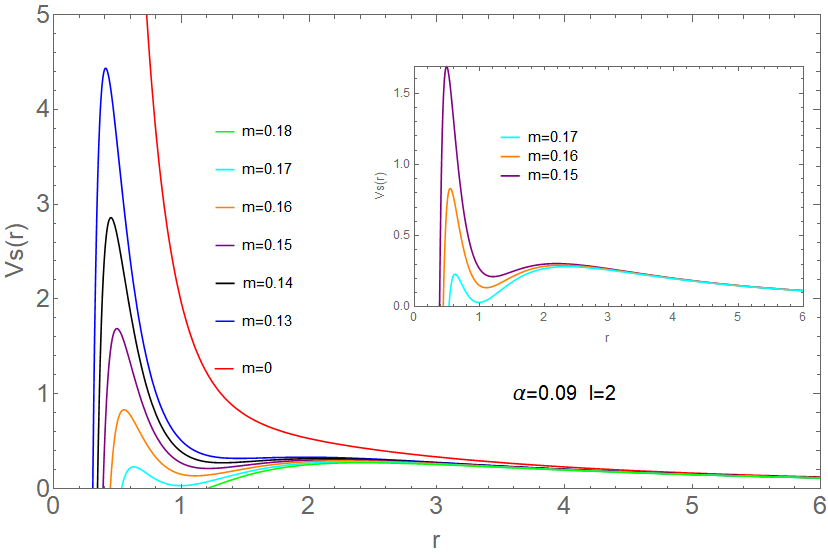}
\caption{ \it Effective scalar potentials $V_s$ for Bardeen black holes. The left and right columns show type-I and type-II solutions, respectively.}
\label{spotencial}
\end{figure}

\begin{figure}[!htbp]
\centering
\includegraphics[width=0.40\textwidth]{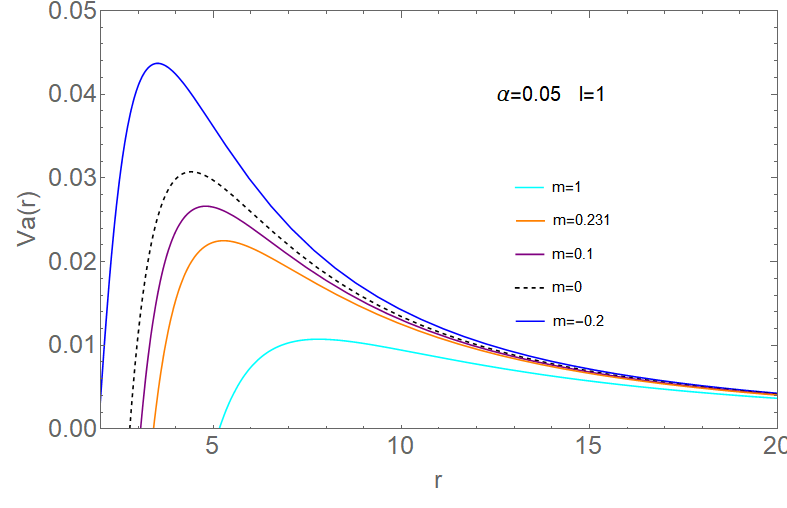} \qquad
\includegraphics[width=0.40\textwidth]{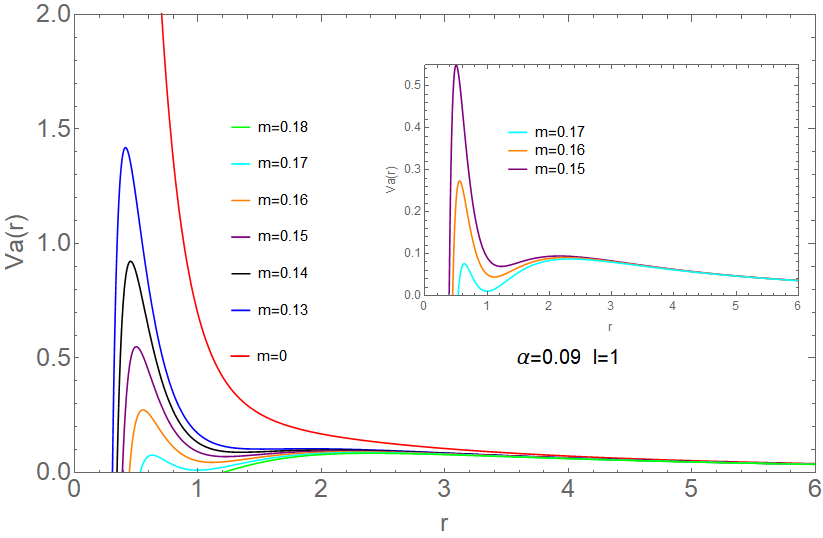}
\includegraphics[width=0.40\textwidth]{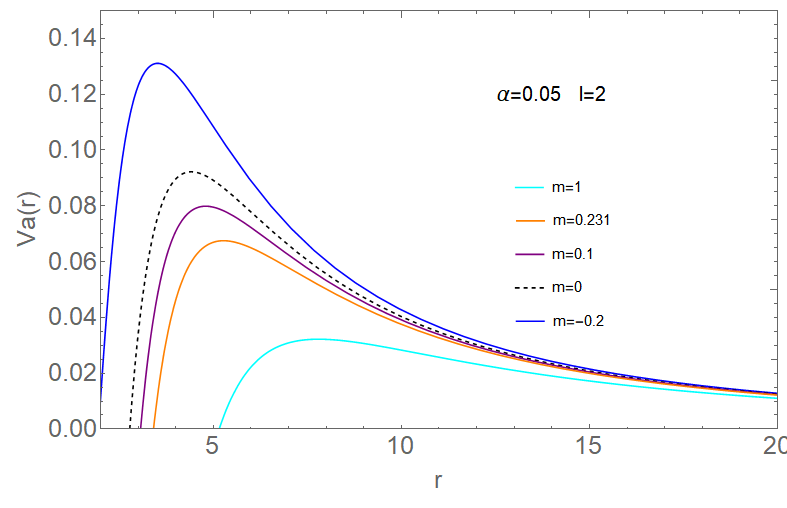} \qquad
\includegraphics[width=0.40\textwidth]{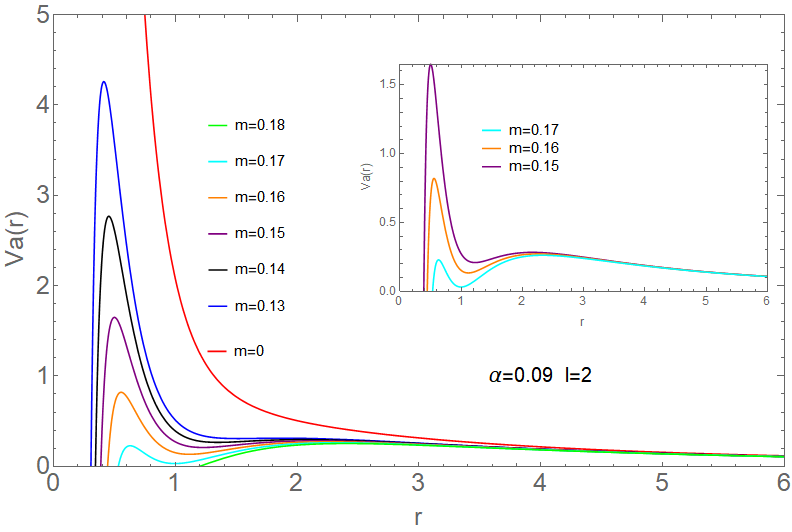}
\caption{ \it Effective potentials $V_a$ of the auxiliary Maxwell field for Bardeen black holes. The left and right columns show type-I and type-II solutions, respectively.}
\label{apotencial}
\end{figure}

The equations of motion for a massless scalar field $\psi$ and an auxiliary Maxwell field ${\cal A}_\mu$ are
\bea\label{boxpsi}
\fft{1}{\sqrt{-g}}\partial_\mu\bigg(\sqrt{-g}g^{\mu\nu}\partial_\nu\psi(t,r,\theta,\varphi)\bigg)=0, \nn \\
\fft{1}{\sqrt{-g}}\partial_\mu\bigg(\sqrt{-g}\,{\cal F}^{\mu\nu}\bigg)=0,
\eea
where ${\cal F}_{\mu\nu}=\partial_\mu{\cal A}_\nu-\partial_\nu{\cal A}_\mu$. For the scalar field, spherical symmetry permits the decomposition
\be
\psi(t,r,\theta,\varphi)=\sum_{\ell, m}\fft {\Phi(t, r)}{r} Y_{\ell m}(\theta, \varphi),
\ee
where $Y_{\ell m}$ are spherical harmonics. The scalar equation in \eqref{boxpsi} then reduces to
\bea
&&-\fft{\partial^2\Phi(t, r)}{\partial t^2}+f^2\fft {\partial^2\Phi(t, r)}{\partial r^2}+ff'\fft{\partial\Phi(t, r)}{\partial r}-V_s(r)\Phi(t, r)=0,
\eea
where a prime denotes a derivative with respect to $r$. Introducing the tortoise coordinate
\be
dr_*=\ft{dr}{f(r)},
\ee
gives the Schr\"odinger-like equation
\be\label{schrodinger}
-\fft{\partial^2\Phi(t,r)}{\partial t^2}+\fft {\partial^2\Phi(t,r)}{\partial r_{*}^2}-V_s(r)\Phi(t,r)=0.
\ee
The scalar effective potential is
\be\label{scalar}
V_{s}=\fft {\ell (\ell+1)}{r^2}f(r)+\fft{f(r)f'(r)}{r}.
\ee
The auxiliary Maxwell field obeys an equation of the same form, with potential
\be\label{max}
V_a=\fft {\ell (\ell+1)}{r^2}f(r).
\ee
The separation of the Maxwell equation and the derivation of Eq.~\eqref{max} are reviewed in Ref.~\cite{Fu:2023drp}.

The effective potentials depend directly on the background geometry. They usually have a single barrier for familiar black-hole solutions, but multiple barriers can occur for wormholes \cite{Liu:2020qia,Peng:2021osd,Su:2024gxp} and for some black holes \cite{Guo:2022umh,Yang:2022ifo,Yang:2024rms,Dong:2020odp}. A wave packet can then be partially trapped between the barriers, producing delayed pulses in the time-domain signal.

Figure~\ref{spotencial} shows the scalar potential for the Bardeen family \eqref{fbf}. In the left column, the $m=0$ member is a regular black hole and the solutions belong to the type-I family. When $m$ is moved away from zero, the black hole becomes singular, but the potential retains a single barrier. Its height decreases as $m$ increases. In the type-II family shown in the right column, the $m=0$ member is a horizonless geometry with a de Sitter core. Among the static members that possess a horizon, the effective potential develops two barriers over a finite range of $m$; their heights again decrease as $m$ increases.

For fixed $m$, the barrier height increases with $\ell$. For $\ell=0$, the scalar potential of a type-II black hole can also contain a negative region. This prevents a stability proof based only on positivity of the potential \cite{Myung:2018vug}, but does not by itself indicate the existence of an unstable mode; a spectral or time-domain analysis is required.

\FloatBarrier
\subsection{Time-domain method and echoes}
\suppressfloats[t]

We next solve the Schr\"odinger-like equation by finite-difference evolution in the time domain. The field is purely ingoing at the event horizon and purely outgoing at spatial infinity. With $r_*$ increasing toward infinity, these conditions read
\bea
&&\text{at the horizon:~~} \partial_{t}\Phi = +\partial_{r_{*}} \Phi,\nn\\
&&\text{at infinity:~~} \partial_{t}\Phi = -\partial_{r_{*}} \Phi .
\eea

On a uniform grid, $t_i=i\Delta t$ and $r_{*,j}=j\Delta r_*$, Eq.~\eqref{schrodinger} becomes

\begin{equation}
\begin{aligned}
0={}&-\frac{\Phi(i+1,j)-2\Phi(i,j)+\Phi(i-1,j)}{\Delta t^2}\\
&+\frac{\Phi(i,j+1)-2\Phi(i,j)+\Phi(i,j-1)}{\Delta r_*^2}
-V(j)\Phi(i,j) .
\end{aligned}
\end{equation}
or, equivalently,
\begin{equation}
\begin{aligned}
\Phi(i+1,j)={}&-\Phi(i-1,j)
+\left(2-2\frac{\Delta t^2}{\Delta r_*^2}-\Delta t^2V(j)\right)\Phi(i,j)\\
&+\frac{\Delta t^2}{\Delta r_*^2}
\left[\Phi(i,j+1)+\Phi(i,j-1)\right] .
\end{aligned}
\end{equation}
We prescribe a stationary Gaussian pulse,
\begin{eqnarray}
&&
\begin{cases}
&\Phi(0,r_{*}) = \exp\!\left[-\fft{(r_{*}-\bar a)^2}{2\sigma^2}\right],\\
&\partial_t\Phi(0,r_{*})=0.
\end{cases}
\end{eqnarray}
\begin{figure}[!htbp]
\centering
\includegraphics[width=0.45\textwidth]{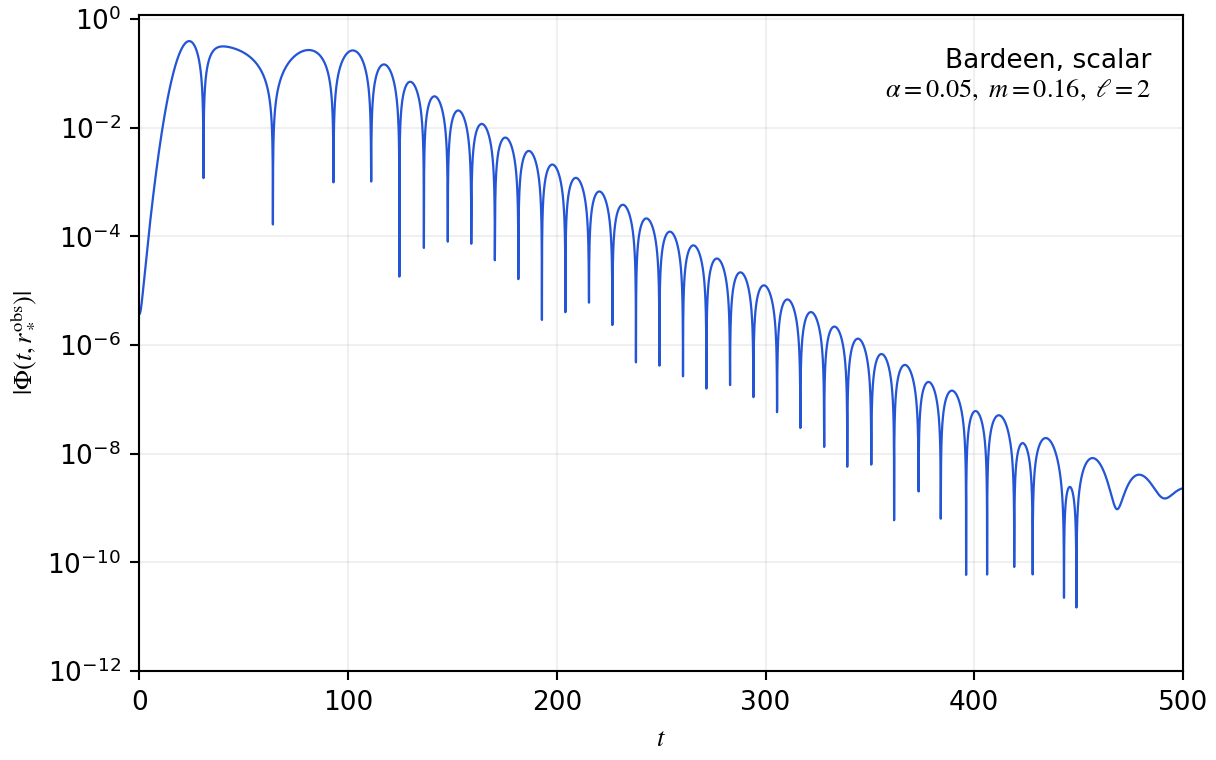} \qquad
\includegraphics[width=0.45\textwidth]{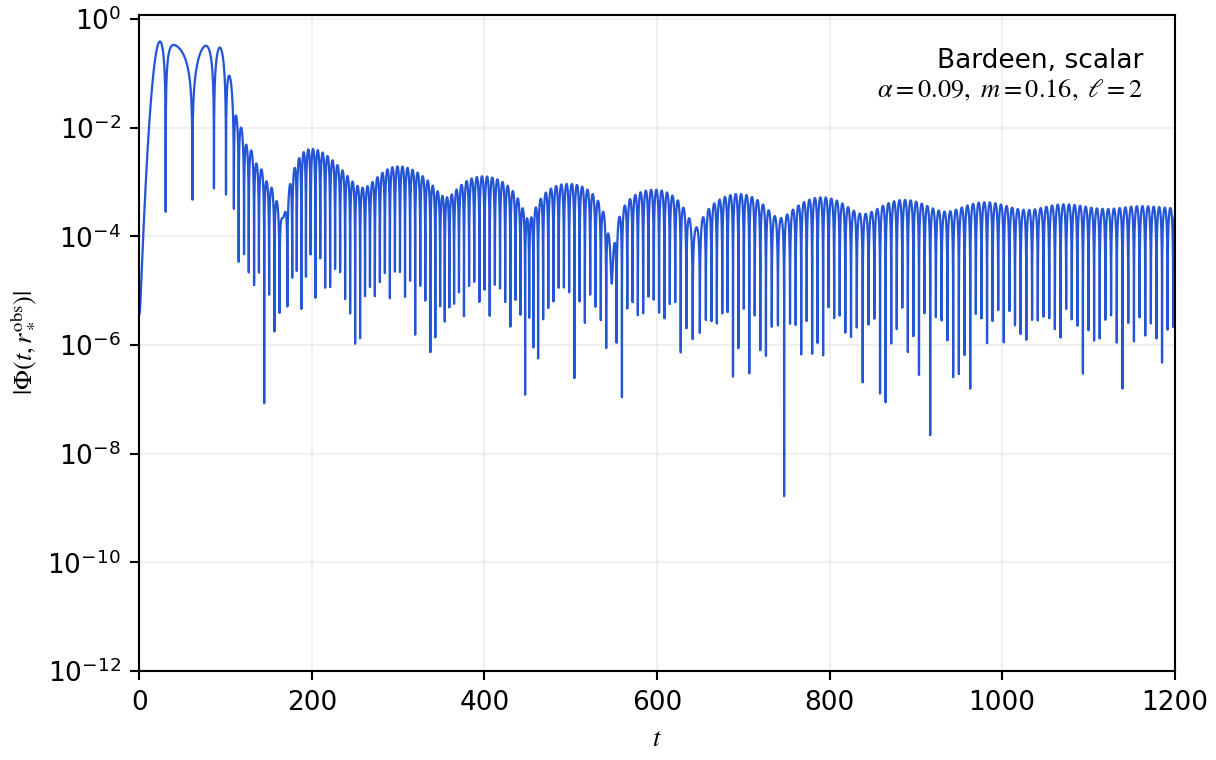} \qquad
\includegraphics[width=0.45\textwidth]{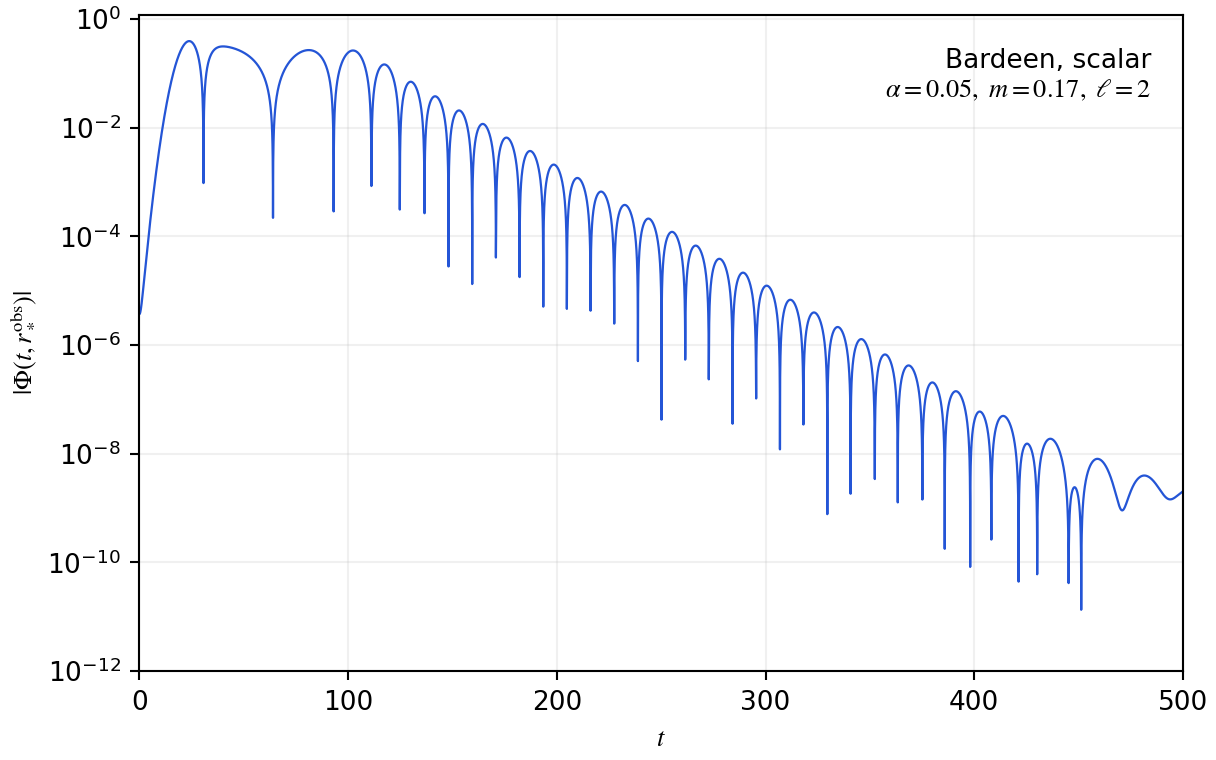} \qquad
\includegraphics[width=0.45\textwidth]{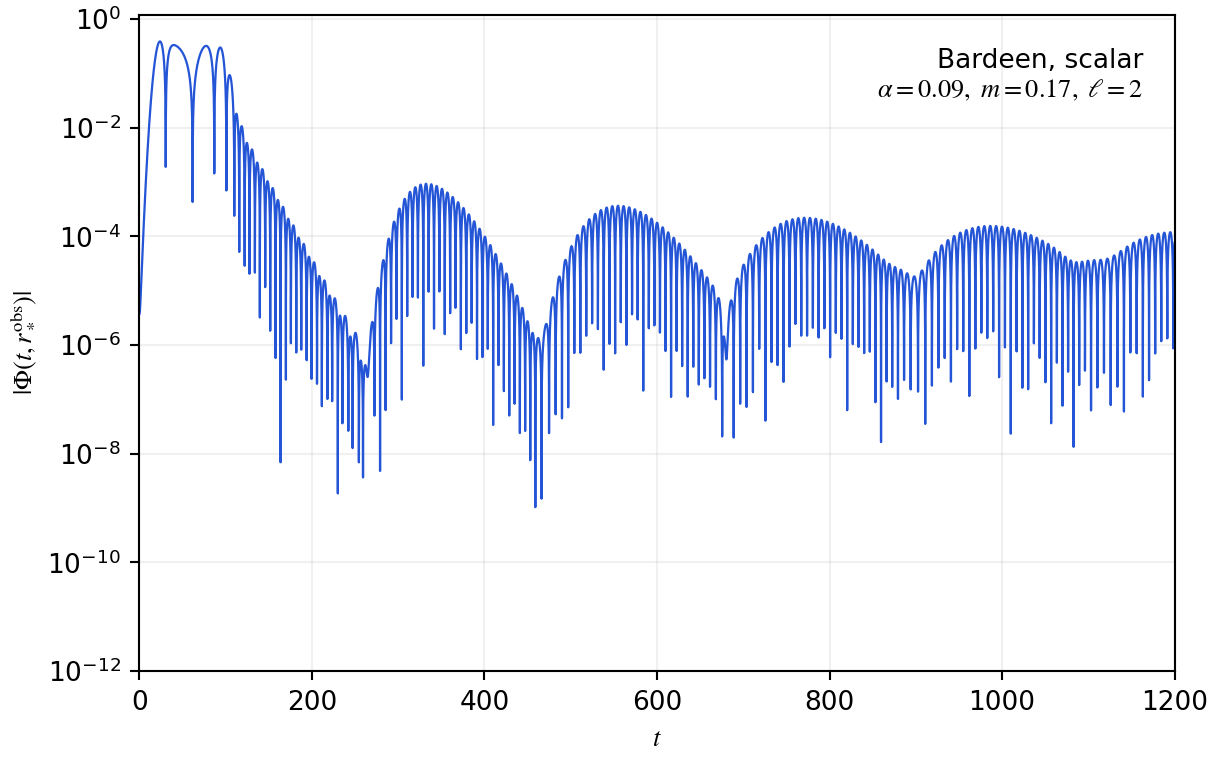} \qquad
\includegraphics[width=0.45\textwidth]{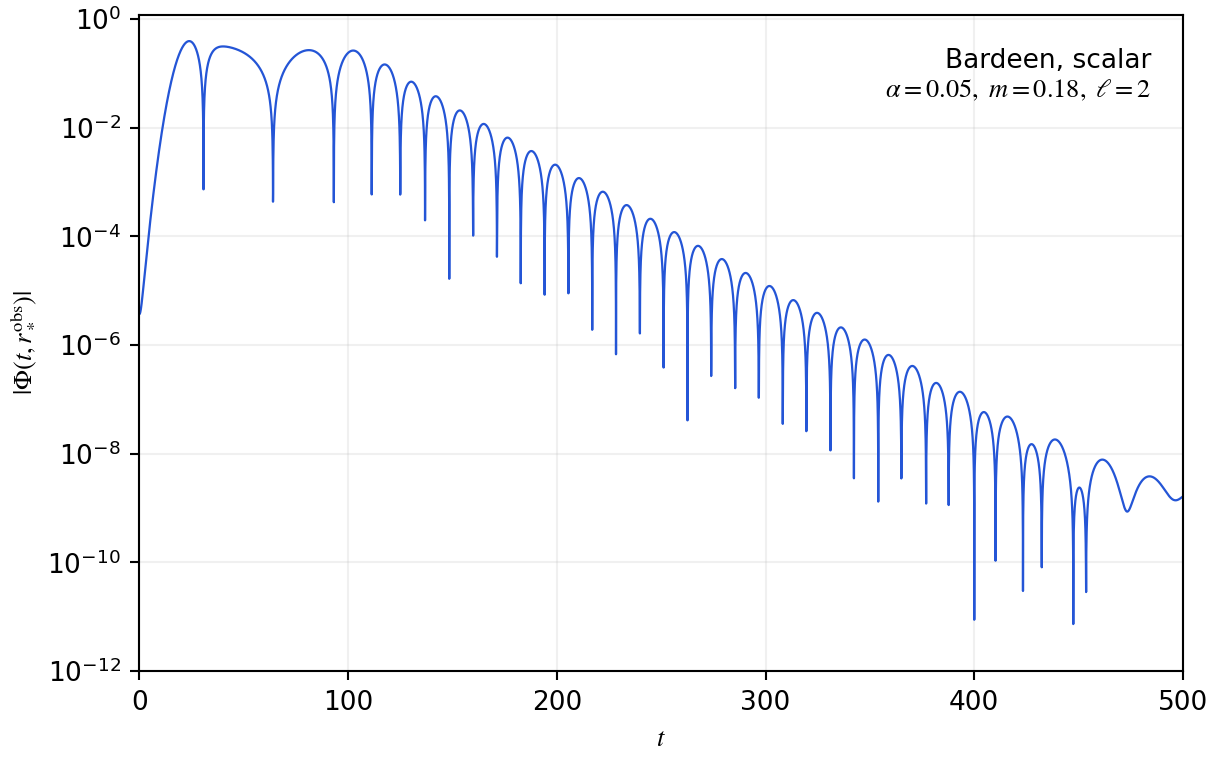} \qquad
\includegraphics[width=0.45\textwidth]{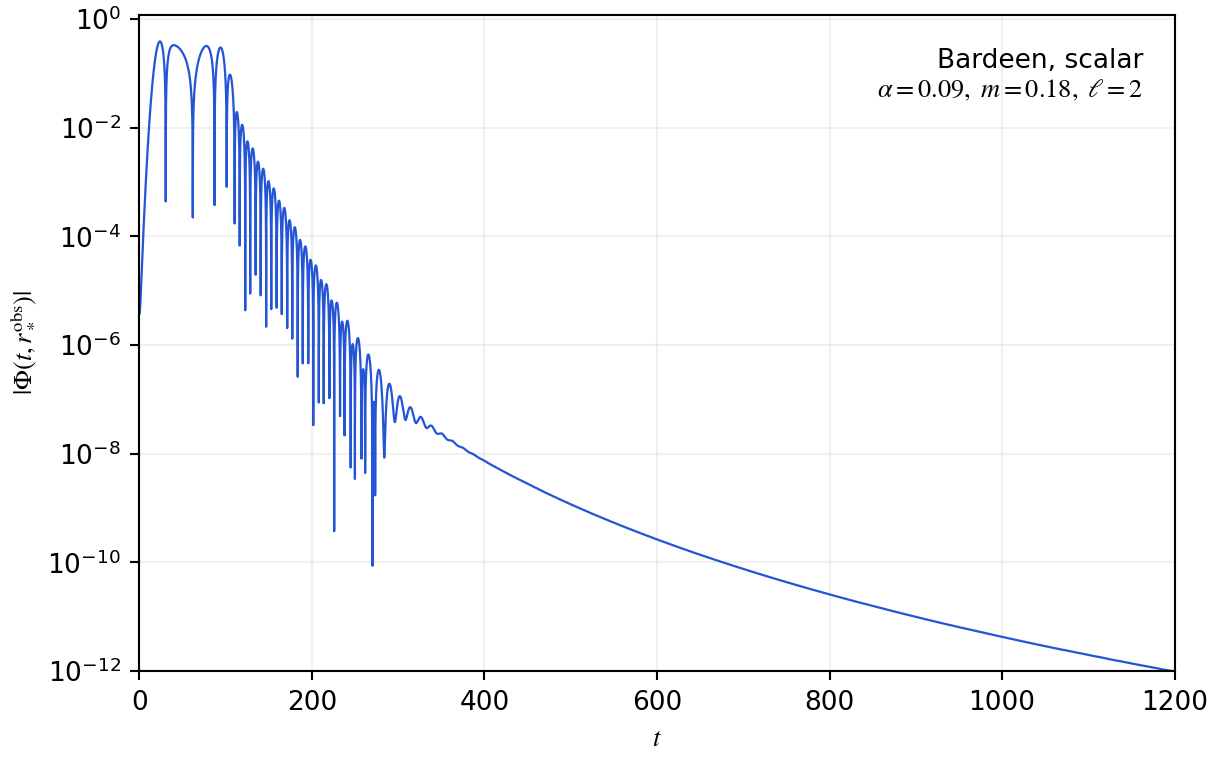} \qquad
\caption{ \it Absolute time-domain amplitude $|\Phi(t,r_*^{\rm obs})|$ on a logarithmic scale for the $\ell=2$ scalar perturbation of Bardeen black holes. The left and right columns show type-I and type-II solutions, respectively.}
\label{sphi}
\end{figure}

\begin{figure}[!htbp]
\centering
\includegraphics[width=0.45\textwidth]{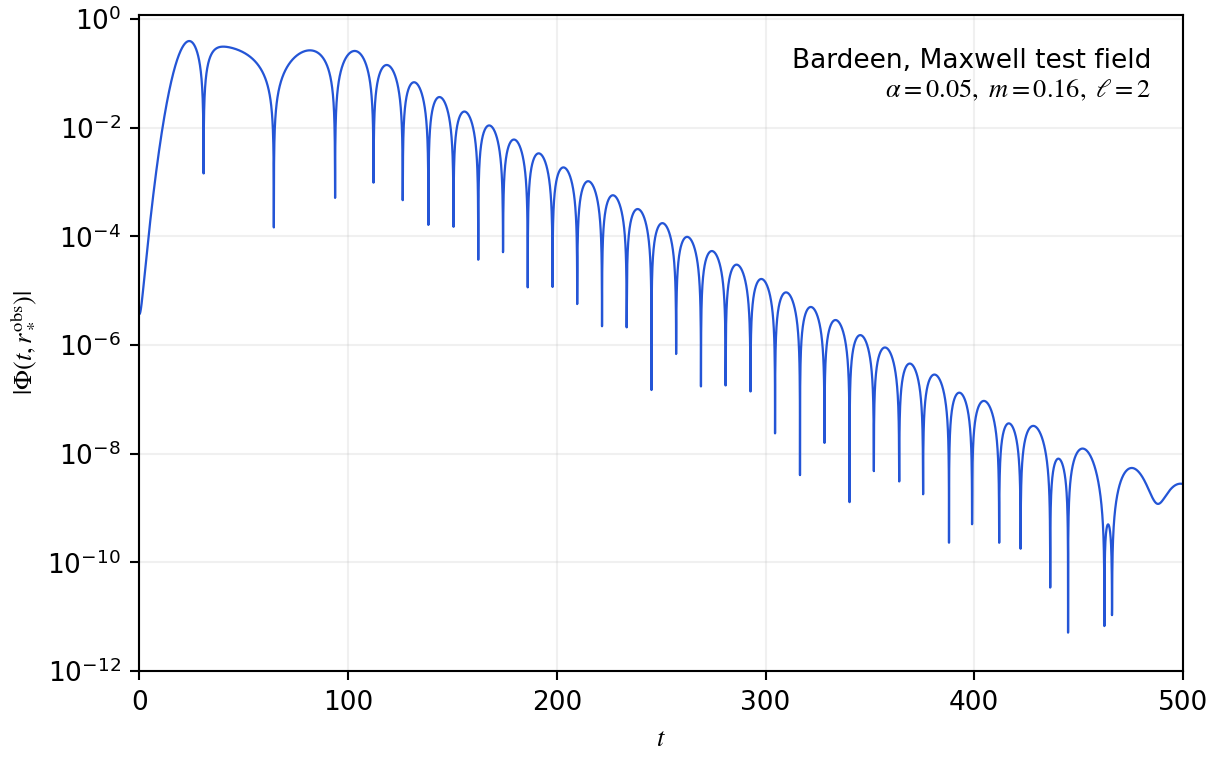} \qquad
\includegraphics[width=0.45\textwidth]{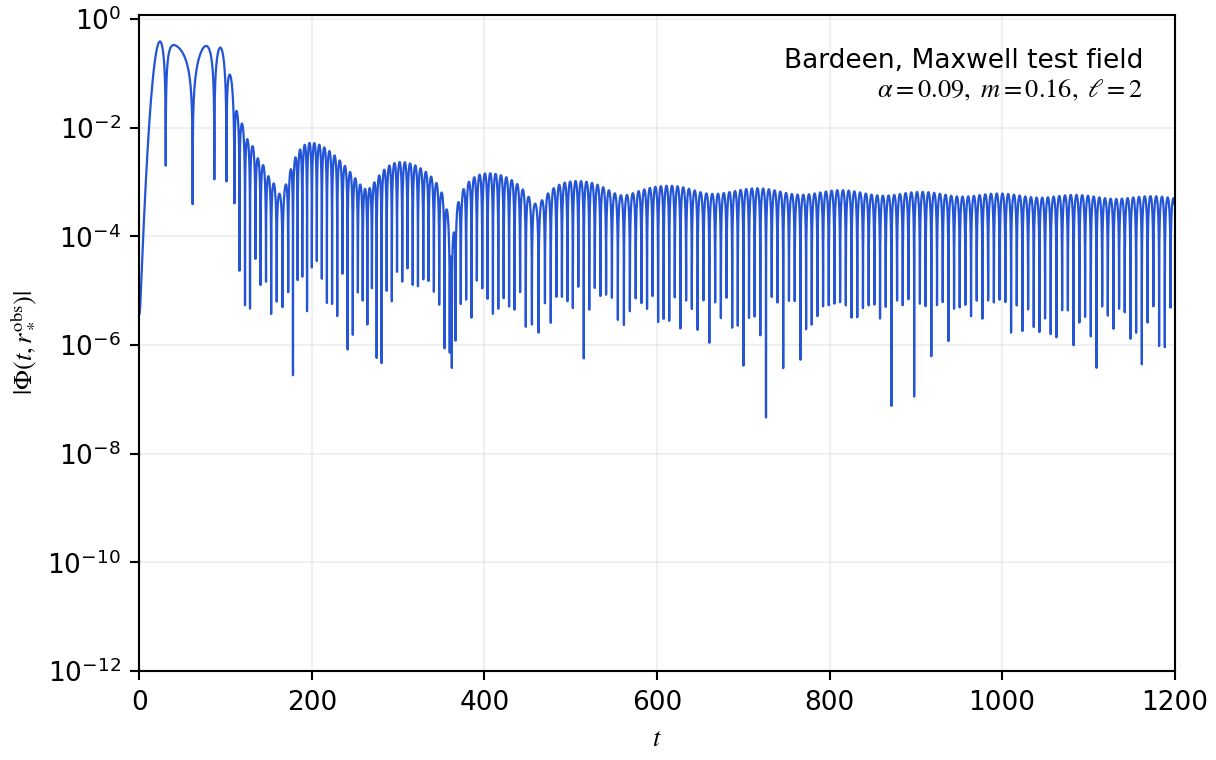} \qquad
\includegraphics[width=0.45\textwidth]{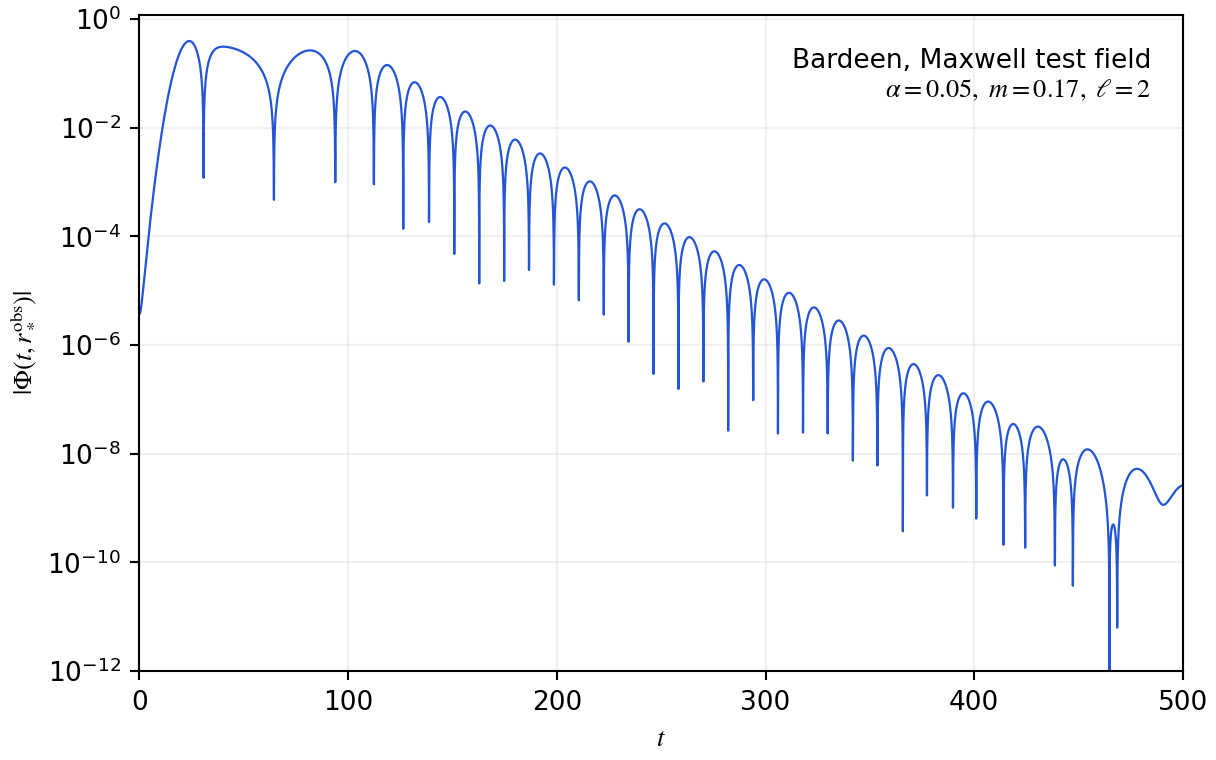} \qquad
\includegraphics[width=0.45\textwidth]{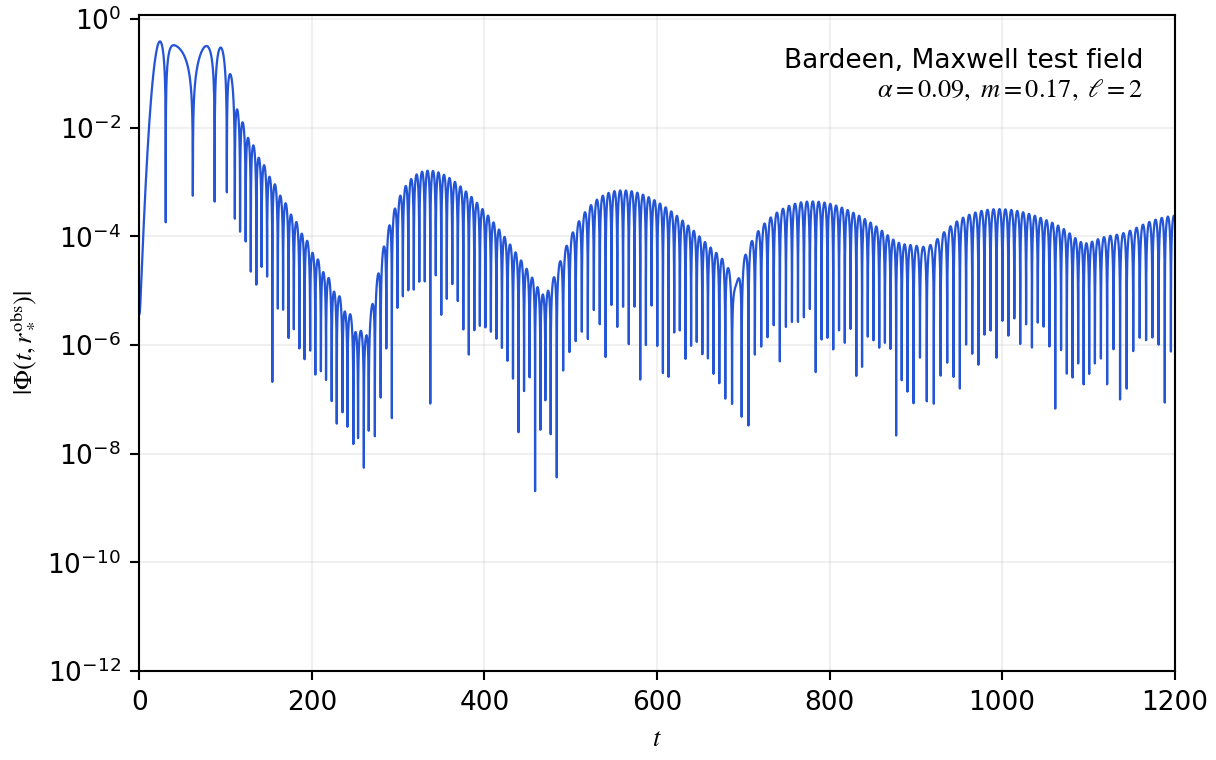} \qquad
\includegraphics[width=0.45\textwidth]{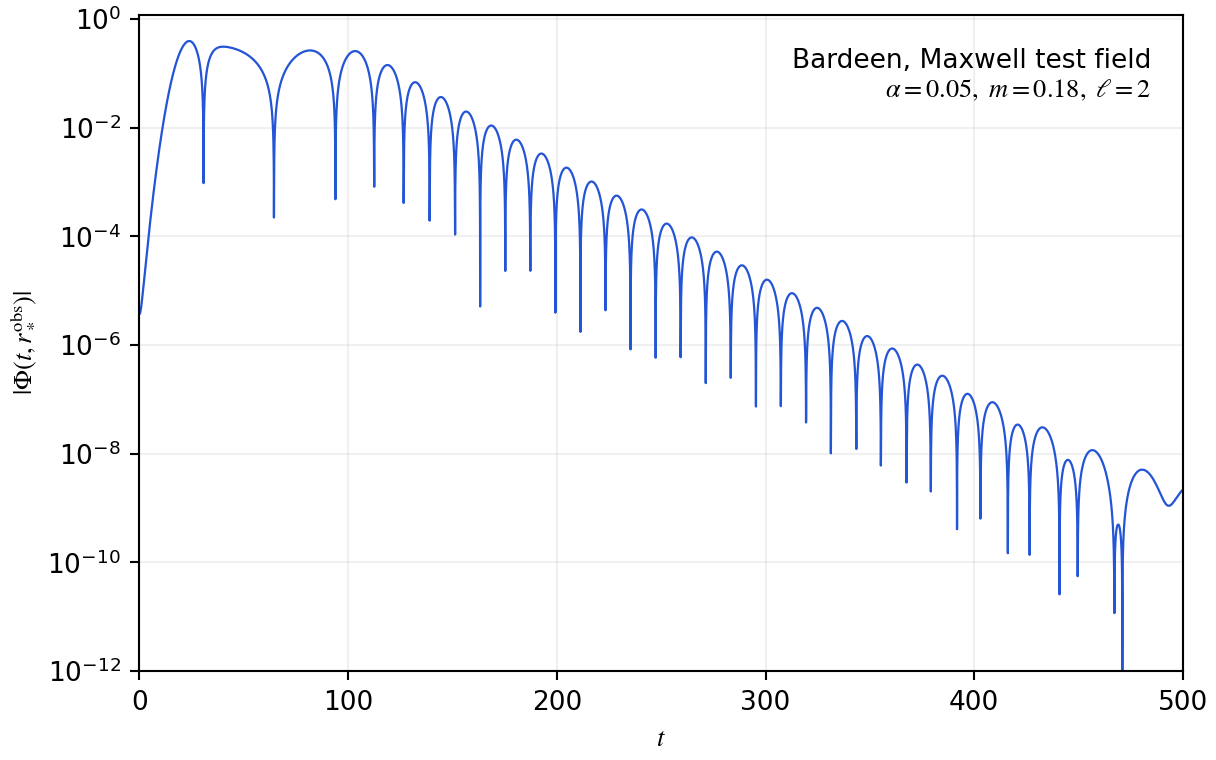} \qquad
\includegraphics[width=0.45\textwidth]{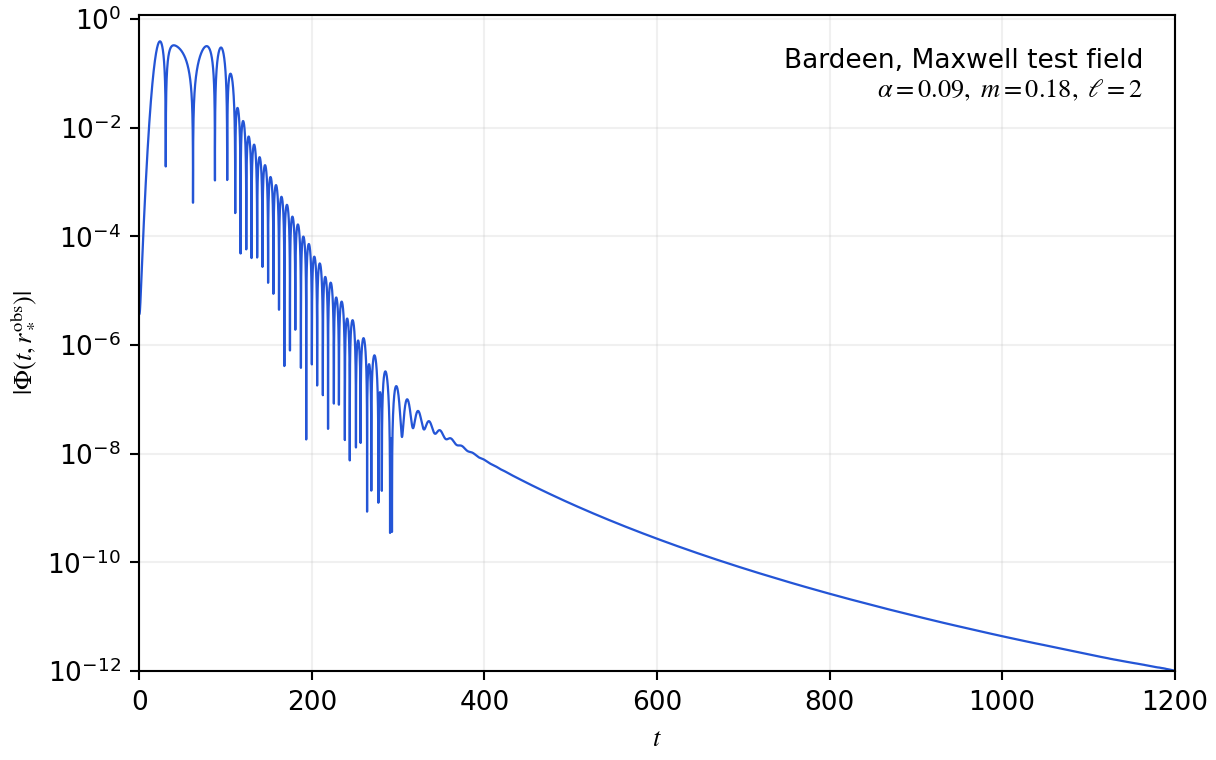} \qquad
\caption{ \it Absolute time-domain amplitude $|\Phi(t,r_*^{\rm obs})|$ on a logarithmic scale for the $\ell=2$ auxiliary Maxwell perturbation of Bardeen black holes. The left and right columns show type-I and type-II solutions, respectively.}
\label{aphi}
\end{figure}

For the profiles shown below, we set $\bar a=35$, $\sigma=5$, $\Delta r_*=0.10$, and $\Delta t=0.05$. The computational domain is $-600\leq r_*\leq700$, the field is recorded at $r_*^{\rm obs}=60$, and the evolution is continued to $t_{\rm max}=1200$. The additive constant in $r_*$ is fixed by placing $r_*=0$ at the outermost maximum of $f(r)/r^2$. Thus the Courant factor is $\ft{\Delta t}{\Delta r_*}=0.5$, which satisfies the von Neumann stability condition for the explicit scheme. The first time step is initialized from the wave equation using the stated vanishing initial time derivative.

The panels in Figs.~\ref{sphi} and \ref{aphi} were regenerated with the documented Julia finite-difference code and show representative evolutions of the scalar and auxiliary Maxwell fields, respectively. A three-grid refinement test gives an observed convergence order of approximately $2.00$, as expected for the second-order scheme.

The left columns of Figs.~\ref{sphi} and \ref{aphi} show type-I solutions for several values of $m$. Their signals are dominated by an ordinary damped ringdown. The right columns show type-II solutions; over an appropriate range of $m$, a sequence of delayed pulses follows the primary ringdown.

Echoes occur for both test fields when the corresponding potential has two barriers. Part of the initial wave packet is temporarily trapped in the intervening cavity and leaks out after repeated reflections. Within the parameter ranges examined here, this double-barrier structure occurs only for type-II black holes.

\FloatBarrier
\subsection{QNMs of the black holes}
\suppressfloats[t]

We now compute the QNM frequencies by fitting the time-domain profiles and by applying the sixth-order WKB approximation. We use the time dependence $e^{-i\omega t}$, so a decaying mode has $\operatorname{Im}\omega<0$. The tabulated modes have $\ell=2$ and overtone number $n=0$. Agreement between the two methods provides a consistency check for configurations in which a single dominant barrier controls the ringdown. A local WKB expansion about one maximum does not, by itself, describe the resonances of a double-barrier cavity or validate the associated echo spectrum.

\begin{table}[!htbp] 
\centering 
\setlength{\tabcolsep}{5.7pt}
\caption{\it Fundamental ($n=0$) QNM frequencies of the $\ell=2$ scalar perturbation for Bardeen black holes.}
\begin{tabular}{ccccc} 
\toprule 
 &$\alpha=0.05$  & &$\alpha=0.2$\\
$m$ & WKB   & Time domain & WKB&Time domain\\
\midrule 
0.2	&0.271917-0.049780i&0.272062-0.049949i&1.598530-0.261658i&1.602447-0.265234i \\

0.4	&0.242223-0.045594i&0.242261-0.045754i &0.695436-0.107823i&0.695440-0.108579i\\

0.6	&0.218902-0.041896i&0.218604-0.041920i&0.511322-0.091316i&0.510838-0.091334i\\
0.8	&0.199951-0.038686i&0.199802-0.038721i&0.413848-0.077714i&0.413574-0.077634i\\

1.0	&0.184174-0.035900i&0.184084-0.035921i
 &0.350057-0.067314i&0.349965-0.067578i

\\

1.2	&0.170799-0.033471i&0.170686-0.033530i
 &0.304208-0.059264i&0.303844-0.059413i

\\

1.4	&0.159293-0.031340i&0.159204-0.031394i
 &0.269372-0.052889i&0.269241 -0.052789i

\\

1.6	&0.149279-0.029457i&0.149207-0.029510i
 &0.241889-0.047732i&0.241571-0.047507i

\\

1.8	&0.140475-0.027784i&0.144676-0.028649i
&0.219598-0.043480i&0.219334-0.043414i

\\

2.0	&0.132670-0.026288i&0.132612-0.026296i
 &0.201128-0.039917i&0.200994-0.039767i

\\
\bottomrule 
\end{tabular}
\label{bqnms}
\end{table}

\begin{table}[!htbp] 
\centering 
\caption{\it Fundamental ($n=0$) QNM frequencies of the $\ell=2$ auxiliary Maxwell perturbation for Bardeen black holes.}
\begin{tabular}{ccccc} 
\toprule 
 &$\alpha=0.05$  & &$\alpha=0.2$\\
$m$ & WKB   & Time domain & WKB&Time domain\\
\midrule 
0.2	&0.258321-0.048943i&0.258179-0.049020i&1.550330-0.259029i& 1.552673-0.264953i
\\
0.4	&0.229832-0.044815i&0.229486-0.044790i&0.665763-0.106804i&0.665023-0.107186i
 \\ 
0.6	&0.207547-0.041171i&0.207277-0.041145i&0.486414-0.089950i&0.486878-0.087021i
 \\ 
0.8	&0.189484-0.038010i&0.189323-0.038015i&0.392761-0.076435i&0.391802-0.077020i
 \\
1.0	&0.174473-0.035268i&0.174363-0.035295i&0.331843-0.066162i&0.331566-0.066085i
 \\
1.2	&0.161761-0.032878i&0.161578-0.032869i&0.288198-0.058229i&0.288297-0.057932i
 \\
1.4	&0.150836-0.030782i&0.150682-0.030778i&0.255099-0.051955i& 0.254816-0.051731i
\\
1.6	&0.141333-0.028932i&0.141210-0.028928i&0.229016-0.046882i& 0.228857-0.047153i
\\
1.8	&0.132983-0.027287i&0.132891-0.027284i&0.207877-0.042702i&0.207536-0.042794i
 \\
2.0	&0.125583-0.025817i&0.125472-0.025832i&0.190370-0.039200i&0.190129-0.039240i
 \\

\bottomrule 
\end{tabular}
\label{bqnma}
\end{table}

\begin{figure}[!htbp]
\centering
\includegraphics[width=0.45\textwidth]{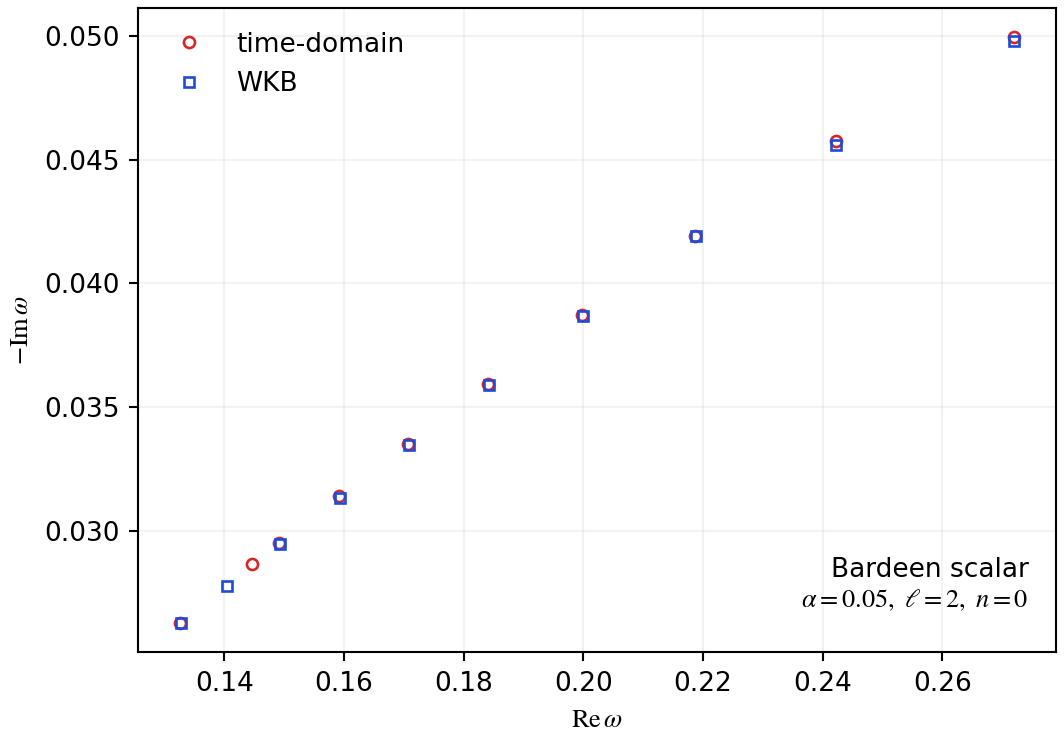} \qquad
\includegraphics[width=0.45\textwidth]{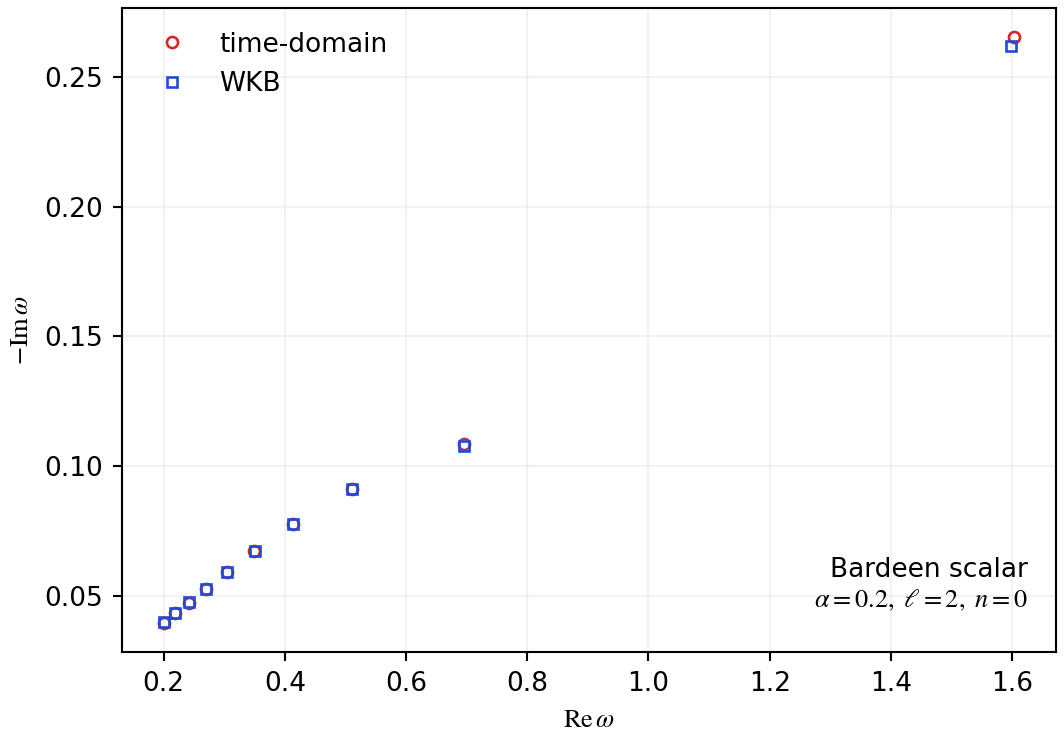} \qquad

\caption{ \it Positive damping rate $-\operatorname{Im}\omega$ versus oscillation frequency $\operatorname{Re}\omega$ for the fundamental scalar modes. Red circles and blue squares denote the time-domain and WKB results, respectively.}
\label{bms}
\end{figure}

\begin{figure}[!htbp]
\centering
\includegraphics[width=0.45\textwidth]{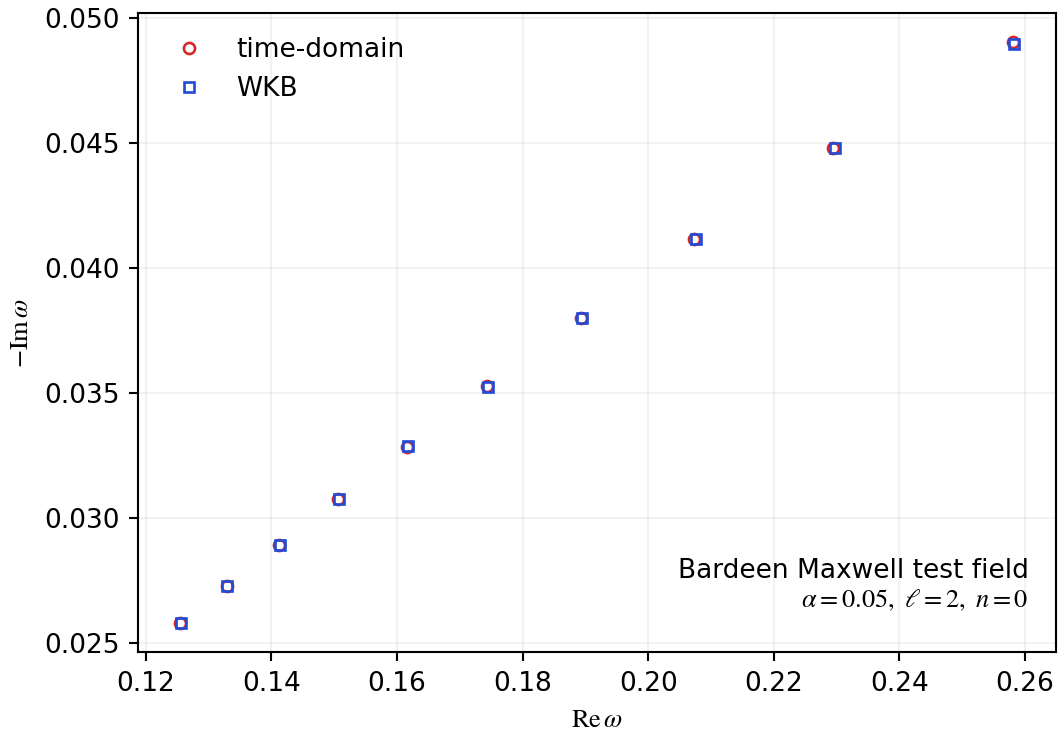} \qquad
\includegraphics[width=0.45\textwidth]{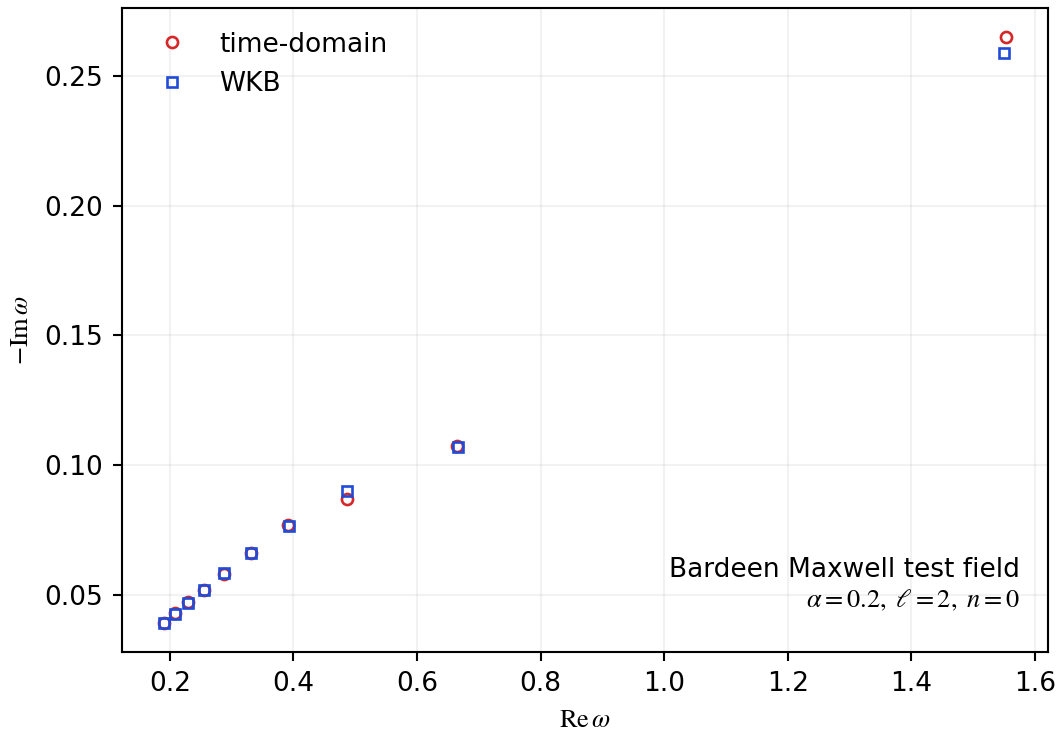} \qquad

\caption{ \it Positive damping rate $-\operatorname{Im}\omega$ versus oscillation frequency $\operatorname{Re}\omega$ for the fundamental auxiliary Maxwell modes. Red circles and blue squares denote the time-domain and WKB results, respectively.}
\label{bma}
\end{figure}

We extract the dominant QNM frequencies from the time-domain profiles using the Prony method \cite{Berti:2007dg}. We also compute them with the same sixth-order WKB approximation following Refs.~\cite{Konoplya:2003ii,Iyer:1986np}. The results are listed in Tables~\ref{bqnms} and \ref{bqnma}.

Table~\ref{bqnms} gives the scalar frequencies, and Table~\ref{bqnma} gives the auxiliary Maxwell frequencies. For all entries, $\operatorname{Im}\omega<0$, so the listed $\ell=2$, $n=0$ probe modes decay in time. This result does not establish stability for other multipoles or overtones, nor does it address the coupled metric--NLED perturbation problem. For the type-I family with $\alpha=0.05$, both $\operatorname{Re}\omega$ and the positive damping rate $-\operatorname{Im}\omega$ decrease as $m$ increases.

The case $\alpha=0.2$ belongs to the type-II family. Here $\operatorname{Re}\omega$ and $-\operatorname{Im}\omega$ decrease rapidly at small $m$ and then vary more slowly as $m$ becomes larger. The same trend appears for the scalar and auxiliary Maxwell fields.

Figures~\ref{bms} and \ref{bma} display the positive damping rate $-\operatorname{Im}\omega$ as a function of $\operatorname{Re}\omega$ and make these trends more apparent. The largest relative differences in the Bardeen tables are about $3.0\%$ in $\operatorname{Re}\omega$ for the scalar mode at $(\alpha,m)=(0.05,1.8)$ and about $3.3\%$ in $-\operatorname{Im}\omega$ for the auxiliary Maxwell mode at $(\alpha,m)=(0.2,0.6)$; the remaining entries generally agree more closely.

\section{Light-ring structure and null geodesics}\label{4}

The double-barrier potentials found above suggest a corresponding structure in the null effective potential. We therefore examine null geodesics of the background metric. The Bardeen family \eqref{fbf} is discussed in this section, and the Hayward results \eqref{hbf} are given in Appendix B.

\FloatBarrier
\subsection{First-order geodesic equation}
\suppressfloats[t]

\begin{figure}[!htbp]
\centering
\includegraphics[width=0.45\textwidth]{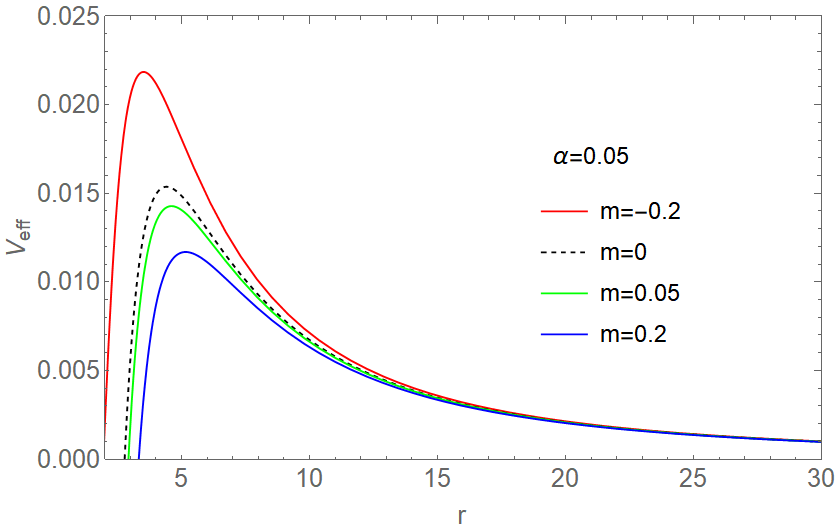} \qquad
\includegraphics[width=0.45\textwidth]{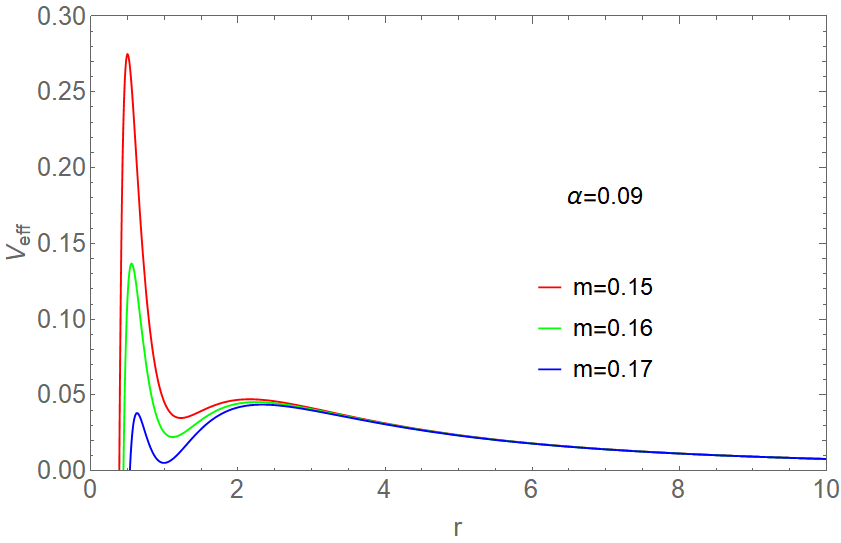}

\caption{ \it Null effective potential $V_{eff}$ for Bardeen black holes. The left and right panels show type-I and type-II solutions, respectively.}
\label{Veffnull}
\end{figure}

The trajectory of a massless test particle is a null geodesic. In the spacetime \eqref{ds}, its worldline can be written as
\be
x^{\mu}(\lambda)=\{t(\lambda),r(\lambda), \theta(\lambda), \varphi(\lambda)\},
\ee
where $\lambda$ is an affine parameter. The tangent vector is
\be
\xi^\mu=\ft{d x^\mu(\lambda)}{d\lambda},
\ee
and satisfies
\be\label{tan}
\xi^{\mu}\xi_\mu=0.
\ee
On the equatorial plane, Eq.~\eqref{tan} becomes
\be
-f(r)(\ft{dt}{d\lambda})^2+f(r)^{-1}(\ft{dr}{d\lambda})^2+r^2 (\ft{d\varphi}{d\lambda})^2=0,
\ee
where we have chosen $\theta=\ft{\pi}{2}$ without loss of generality. The conserved energy and angular momentum are
\be
E=f(r)\ft{dt}{d\lambda}, \qquad L=r^2\ft{d\varphi}{d\lambda},
\ee
along each geodesic. The first-order orbital equation is therefore
\be\label{neo}
\left(\ft{dr}{d\varphi}\right)^2=r^4\left(\ft{1}{b^2}-V_{eff}\right).
\ee
Here the impact parameter $b$ and effective potential $V_{eff}$ are
\be
b=\ft{L}{E}, \qquad V_{eff}=\ft{f(r)}{r^2}.
\ee
Circular null orbits satisfy $V_{eff}'(r_c)=0$. A local maximum, $V_{eff}''(r_c)<0$, is unstable, whereas a local minimum, $V_{eff}''(r_c)>0$, is stable. In the terminology used below, a light ring refers to an unstable circular null orbit. Each unstable orbit at $r_{LR,k}$ has a critical impact parameter
\be
\ft{1}{b_{c,k}^2}=V_{eff}(r_{LR,k}).
\ee
Representative potentials are shown in Fig.~\ref{Veffnull}. For type-I black holes, the effective potential has one maximum throughout the range of $m$ and hence one unstable light ring. For type-II black holes, it can develop two maxima over a finite range of $m$. The two unstable light rings are necessarily separated by a local minimum of $V_{eff}$, which corresponds to a stable circular null orbit.

For example, at $(\alpha,m)=(0.09,0.16)$ the Bardeen solution has an event horizon at $r_h\simeq0.44317$. The three exterior extrema occur at $r\simeq0.55449$ (unstable), $1.11437$ (stable), and $2.25009$ (unstable). These values provide a direct numerical check of the multiple-trapping configuration displayed below.

The null effective potential is proportional to the auxiliary Maxwell potential in Eq.~\eqref{max}, and the scalar potential \eqref{scalar} has the same leading form in the eikonal limit $\ell\to\infty$. This relation concerns null geodesics of the background metric. In NLED, physical electromagnetic rays generally follow null geodesics of an effective optical metric \cite{Novello:1999pg}, so the trajectories computed here should not be identified directly with an observable photon shadow.

\FloatBarrier
\subsection{Null geodesics}
\suppressfloats[t]

\begin{figure}[!htbp]
\centering
\includegraphics[width=0.45\textwidth]{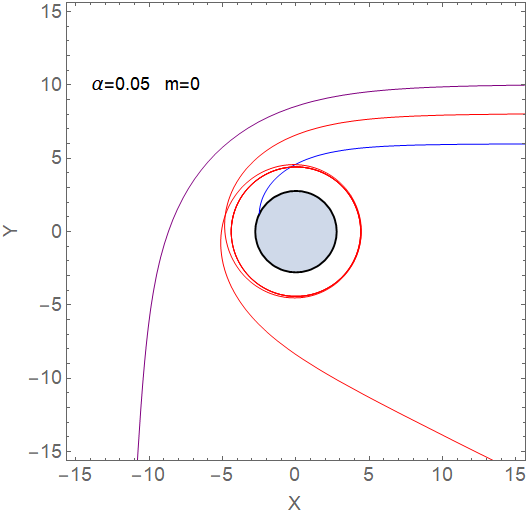} \qquad
\includegraphics[width=0.44\textwidth]{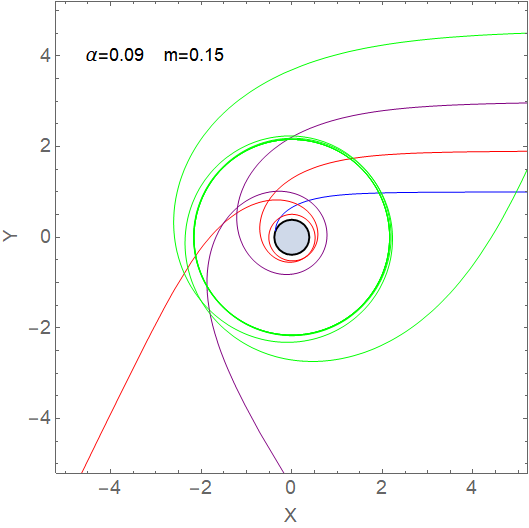}

\caption{ \it Representative null geodesics of the Bardeen black holes. The left panel shows a type-I black hole with one unstable light ring. The right panel shows a type-II black hole with two unstable light rings and an intervening stable circular null orbit.}
\label{geop}
\end{figure}

We integrate Eq.~\eqref{neo} numerically for rays incident from spatial infinity with fixed impact parameter $b$. Representative trajectories are shown in Fig.~\ref{geop} in the Cartesian coordinates
\be
X=r \cos(\varphi), \qquad Y=r \sin(\varphi).
\ee

The left panel of Fig.~\ref{geop} shows a black hole with one unstable light ring. A ray with $b=b_c$ asymptotically approaches the circular orbit (red curve), whereas a ray with $b<b_c$ crosses it and falls through the horizon (blue curve). The right panel shows a black hole with two unstable light rings, each with its own critical impact parameter. In this case, some rays cross the outer light ring, turn around in the inner region, and escape to infinity (purple curve).

This additional scattering branch is accessible from infinity only when the inner maximum of $V_{eff}$ is higher than the outer one. If the inner maximum is lower, a ray that crosses the outer barrier cannot be reflected back through it and instead falls into the black hole.

\begin{figure}[!htbp]
\centering
\includegraphics[width=0.45\textwidth]{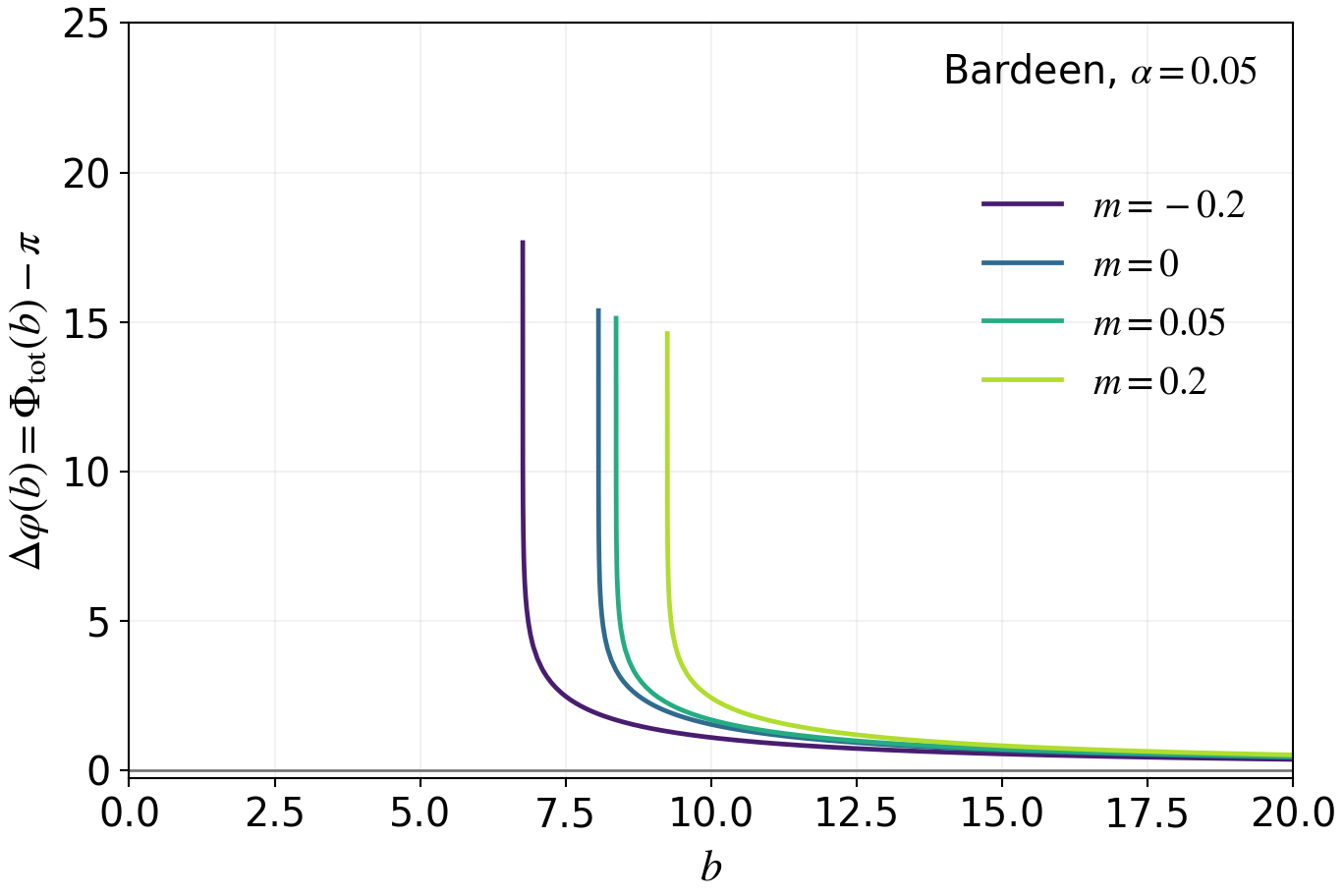} \qquad
\includegraphics[width=0.45\textwidth]{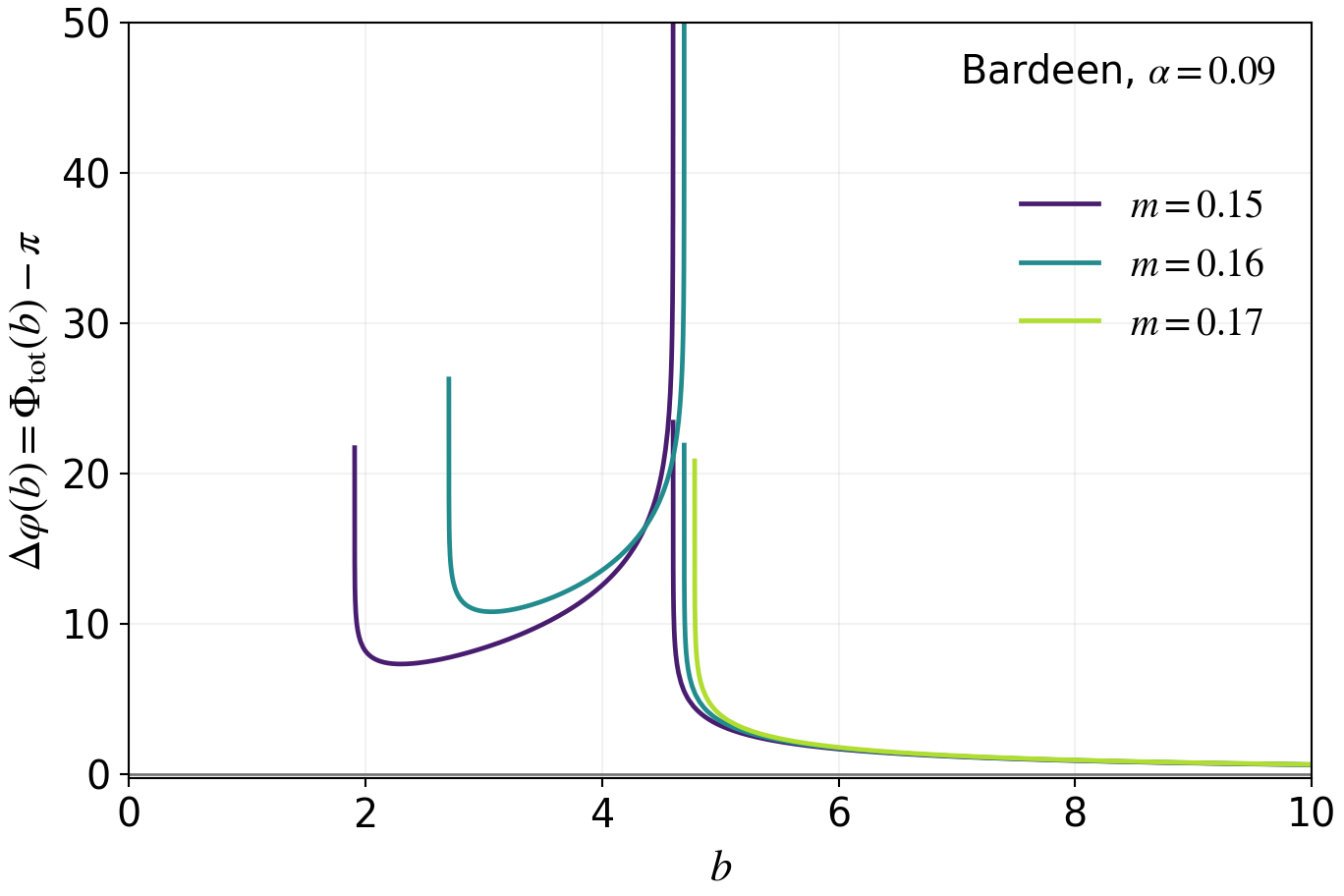}

\caption{ \it Deflection angle of returning null rays for Bardeen black holes with $\alpha=0.05$ (left) and $0.09$ (right). Separate curve segments are the connected scattering branches delimited by accessible critical impact parameters; blank low-$b$ intervals denote capture.}
\label{tot}
\end{figure}

For a ray that returns to infinity, let $r_t$ denote its radial turning point. The total accumulated azimuth and the deflection angle are
\be
\Phi_{\rm tot}(b)=2\int_{r_t}^{\infty}\frac{dr}{r^2\sqrt{\ft{1}{b^2}-V_{eff}(r)}},
\qquad
\Delta\varphi(b)=\Phi_{\rm tot}(b)-\pi .
\ee
Captured rays have no radial turning point and are therefore not assigned a scattering deflection angle by this formula. Figure~\ref{tot} shows the deflection angle for returning rays. The left and right panels correspond to black holes with one and two unstable light rings, respectively. At an accessible unstable orbit, $\Delta\varphi$ diverges logarithmically; the connected returning-ray intervals are therefore plotted separately rather than joined across the critical value of $b$.

\section{Conclusion and outlook}\label{5}

Regular black holes in GR can occur as special members of broader solution families. Using the Bardeen and Hayward families as examples, we studied massless scalar and auxiliary Maxwell perturbations of their regular members and singular counterparts. Depending on the coupling of the theory, the regular member is either a black hole (type I) or a horizonless geometry with a de Sitter core (type II). The type-I solutions retain a single-barrier potential as $m$ varies and exhibit an ordinary damped ringdown. In part of the type-II branch, the potential develops two barriers and the time-domain signal contains late-time echoes. For the single-barrier configurations considered in the frequency tables, the WKB frequencies agree with those extracted from the time-domain profiles. The decay of the listed $\ell=2$, $n=0$ probe modes is not a general stability result: other multipoles and overtones, as well as coupled metric--NLED perturbations, remain to be analyzed.

We also studied null geodesics of the background metric. In the same part of the type-II branch, the null effective potential develops two maxima separated by a minimum, corresponding to two unstable light rings and an intervening stable circular null orbit. This structure is consistent with the double barriers responsible for the echoes. Within the Bardeen and Hayward families, echoes and multiple circular null orbits distinguish this subset of the singular branch from the regular member. They do not, however, provide a model-independent test for the presence of a central singularity. Moreover, because physical photons in NLED propagate in an effective optical metric, our background-metric geodesics should not be interpreted directly as a prediction for the observed shadow.

Reference~\cite{Huang:2025uhv} showed that analogous singular branches also arise for other matter models minimally coupled to GR. It would be useful to determine whether double-barrier potentials and echoes persist in those examples. Further work should also address coupled gravitational--matter perturbations and, for NLED backgrounds, null rays of the effective optical metric.

\section*{Appendix A: QNMs of the Hayward class}\label{app}

The Hayward family has the same qualitative features as the Bardeen family. For the $m=0$ Hayward seed, the extremal coupling is
\be
\alpha_H^{\rm ext}=\frac{2^{2/3}}{18},
\qquad r_h=2^{1/3}.
\ee
The seed is a two-horizon regular black hole for $\alpha<\alpha_H^{\rm ext}$, an extremal regular black hole at equality, and a horizonless geometry with a de Sitter core for $\alpha>\alpha_H^{\rm ext}$. This appendix presents the corresponding QNM results.

\begin{figure}[!htbp]
\centering
\includegraphics[width=0.45\textwidth]{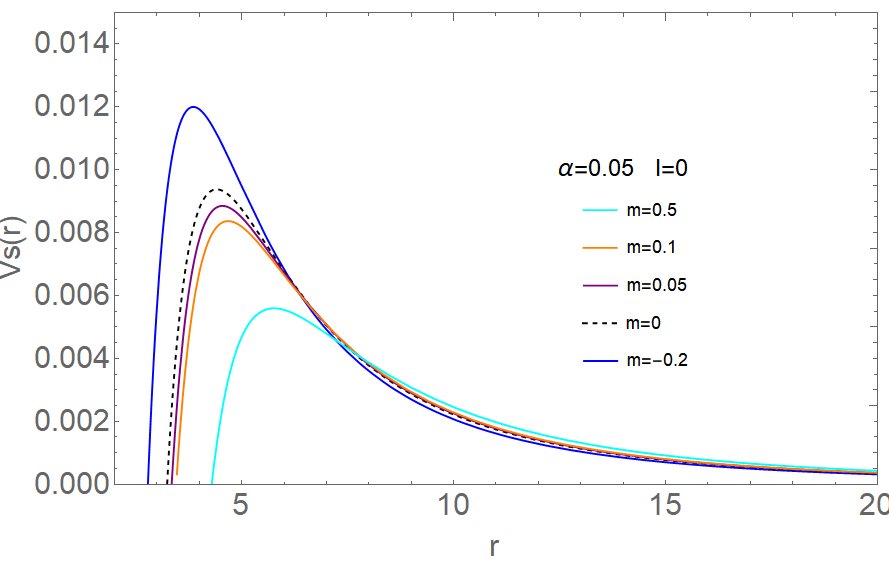}
\includegraphics[width=0.45\textwidth]{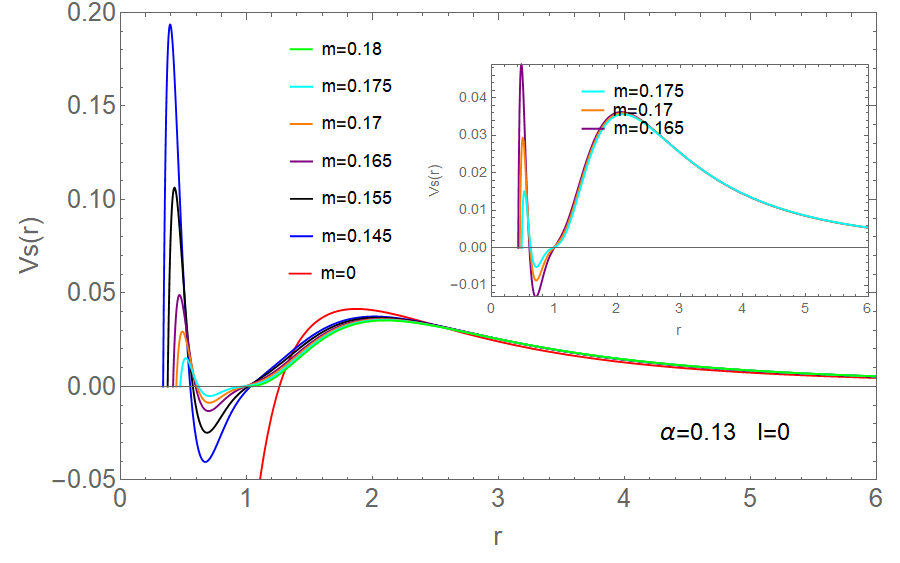} 
\includegraphics[width=0.45\textwidth]{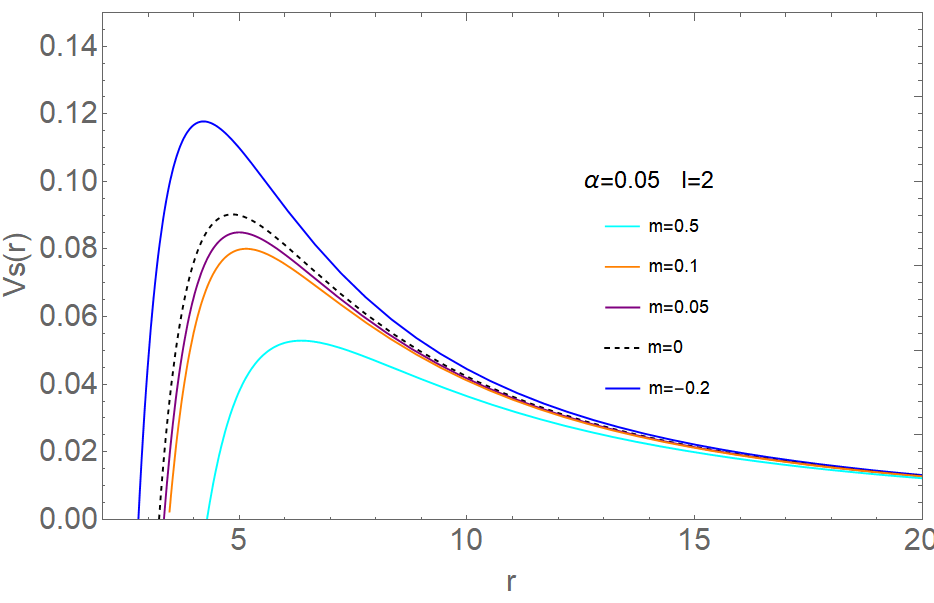}
\includegraphics[width=0.45\textwidth]{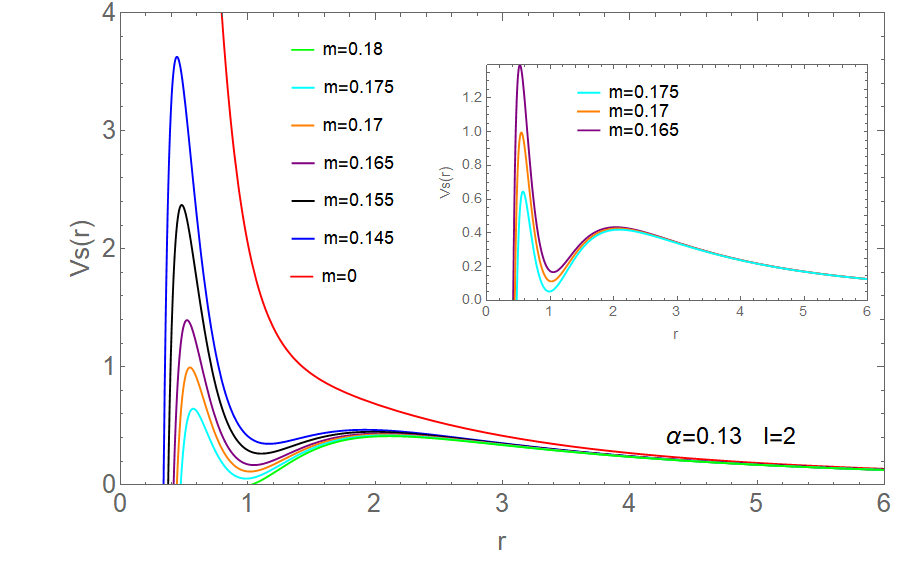} 
\caption{ \it Effective scalar potentials $V_s$ for Hayward black holes. The left and right columns show type-I and type-II solutions, respectively.}
\label{Hsv}
\end{figure}

\begin{figure}[!htbp]
\centering 
\includegraphics[width=0.45\textwidth]{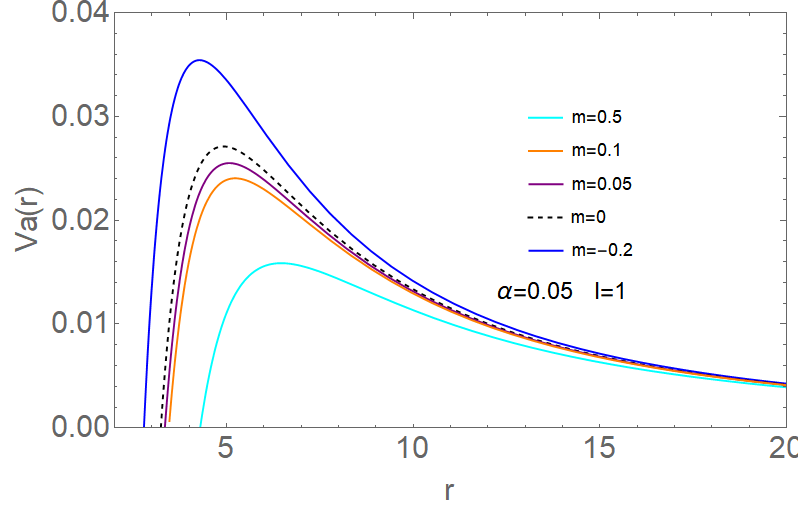}
\includegraphics[width=0.45\textwidth]{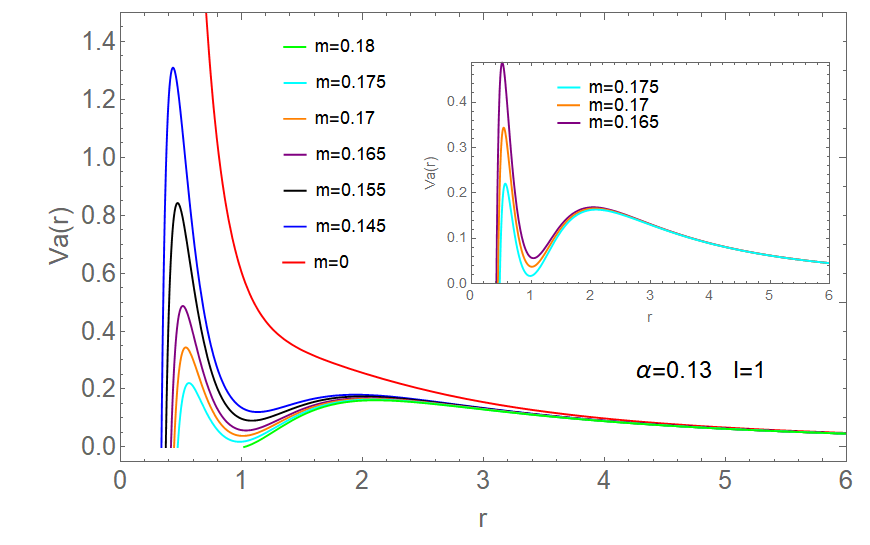} 
\includegraphics[width=0.45\textwidth]{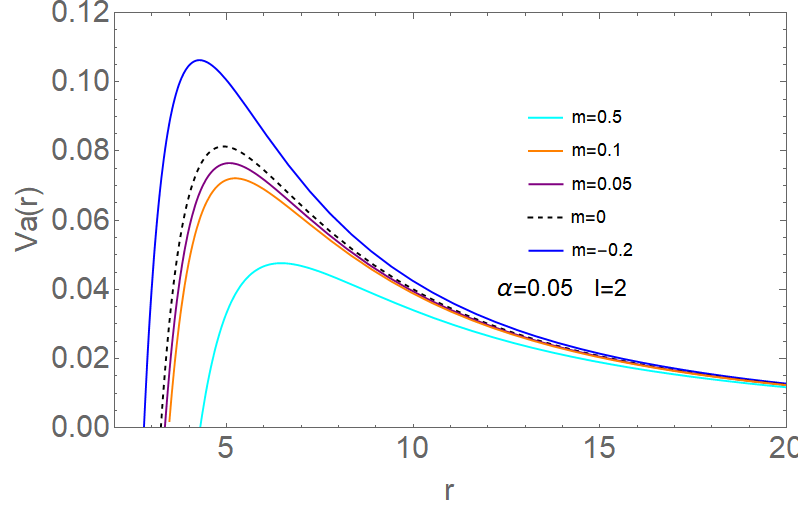}
\includegraphics[width=0.45\textwidth]{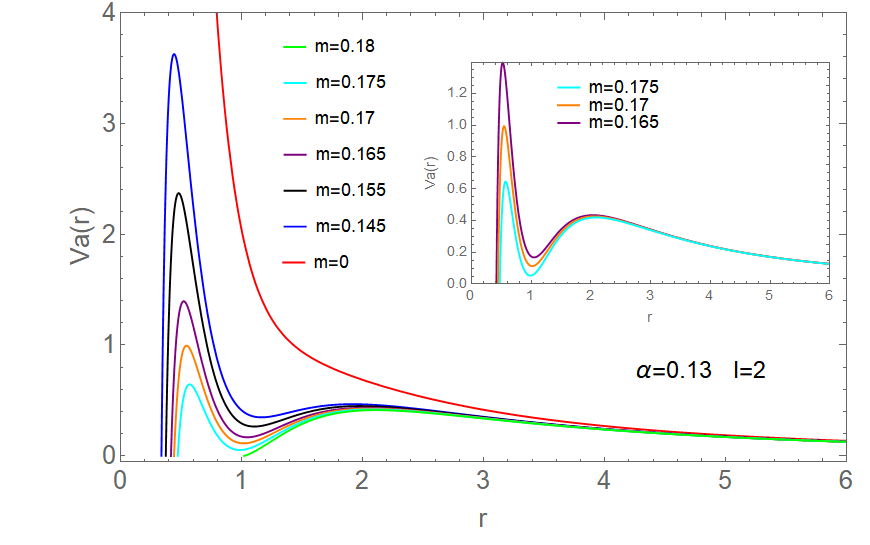} 
\caption{ \it Effective potentials $V_a$ of the auxiliary Maxwell field for Hayward black holes. The left and right columns show type-I and type-II solutions, respectively.}
\label{Hav}
\end{figure}

\begin{figure}[!htbp]
\centering
\includegraphics[width=0.45\textwidth]{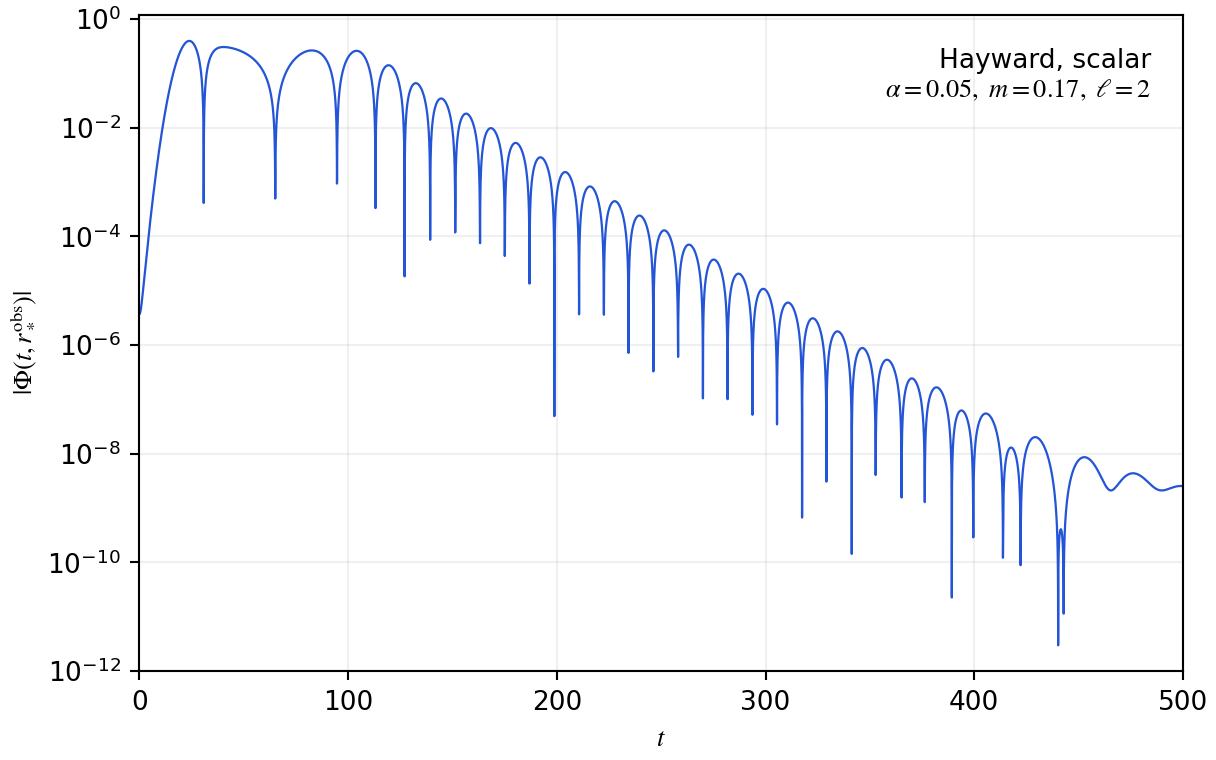} \qquad
\includegraphics[width=0.45\textwidth]{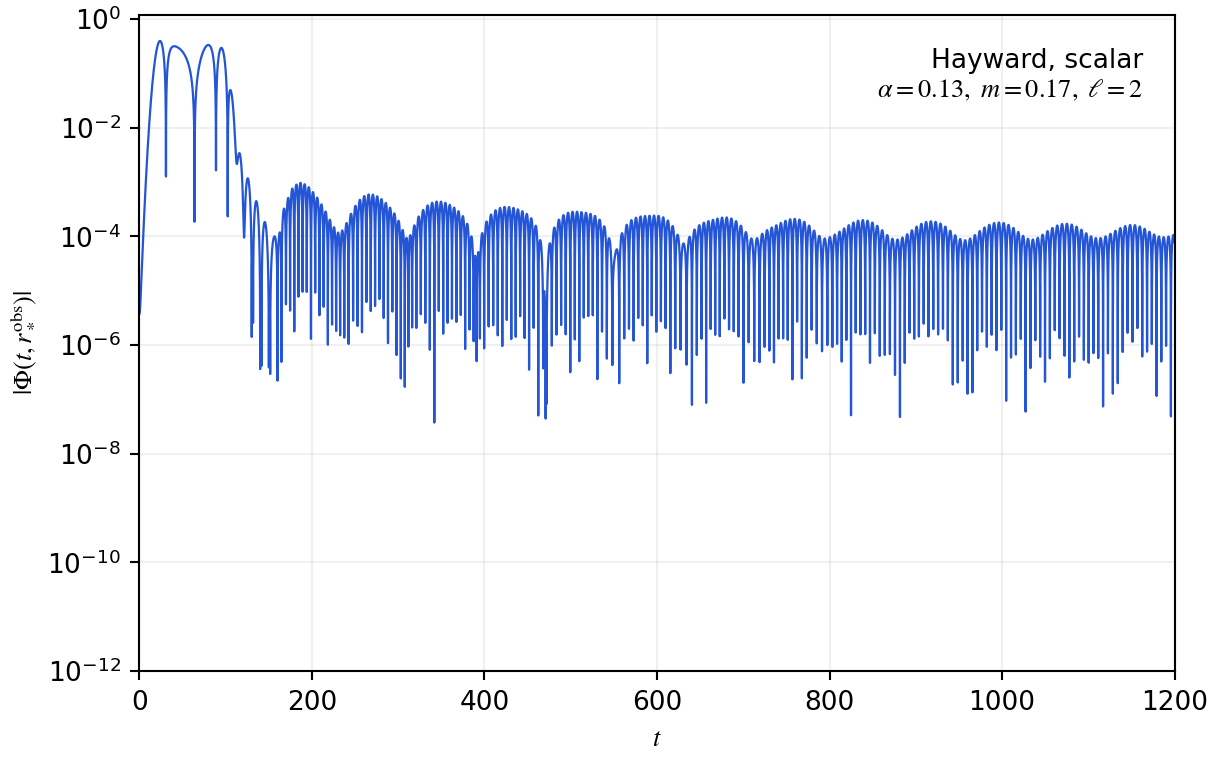} \qquad

\includegraphics[width=0.45\textwidth]{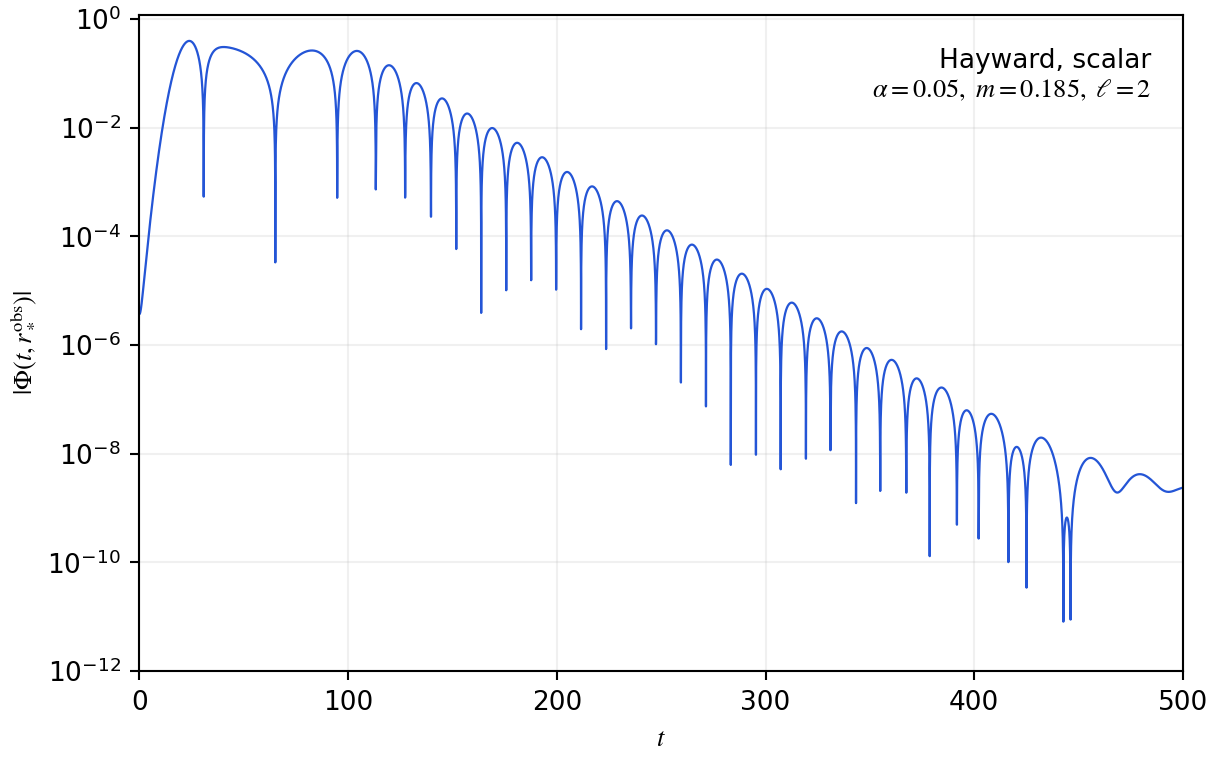} \qquad
\includegraphics[width=0.45\textwidth]{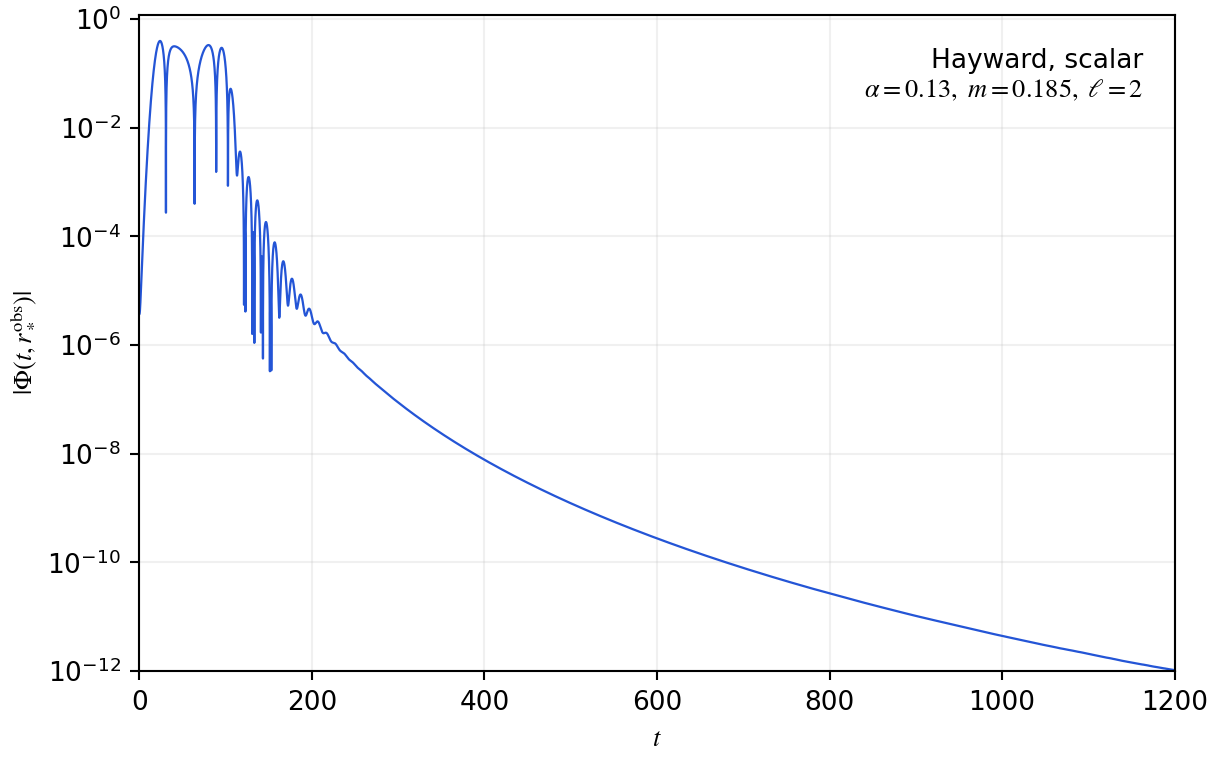} \qquad
\caption{ \it Absolute time-domain amplitude $|\Phi(t,r_*^{\rm obs})|$ on a logarithmic scale for the $\ell=2$ scalar perturbation of Hayward black holes. The left and right columns show type-I and type-II solutions, respectively.}
\label{Hsphi}
\end{figure}

\begin{figure}[!htbp]
\centering

\includegraphics[width=0.45\textwidth]{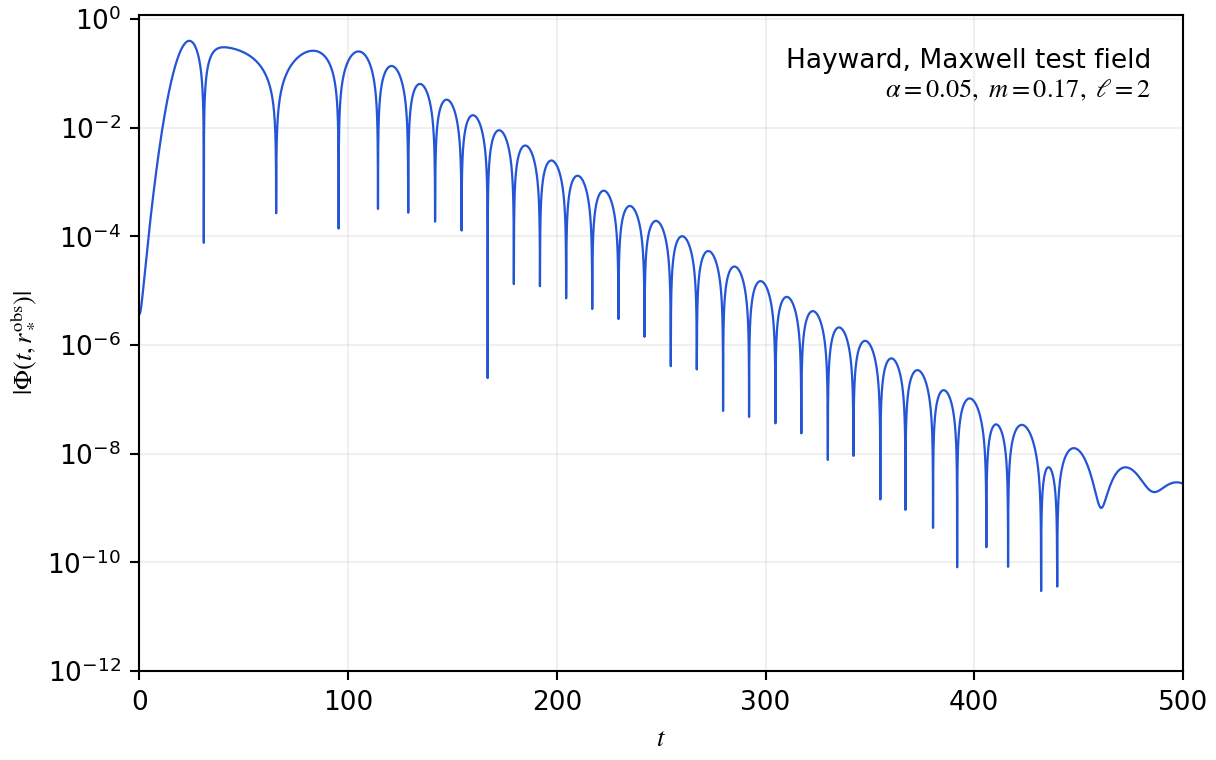} \qquad
\includegraphics[width=0.45\textwidth]{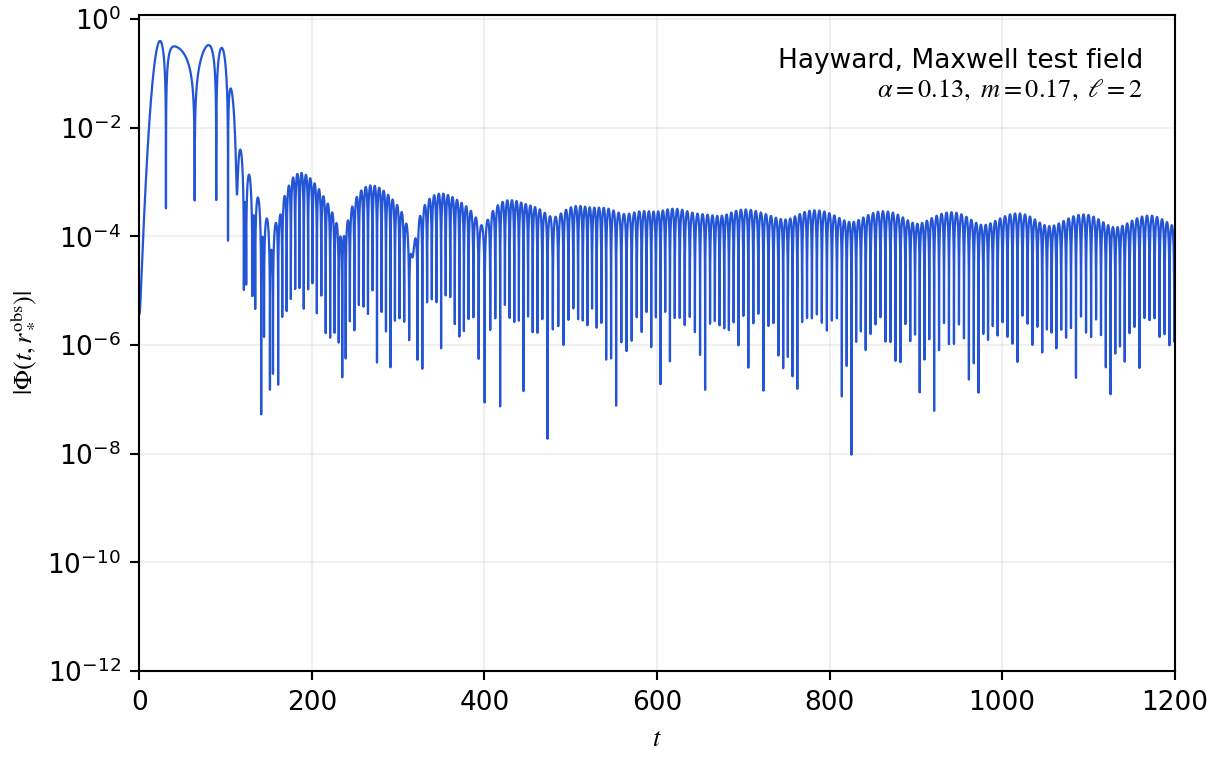} \qquad
\includegraphics[width=0.45\textwidth]{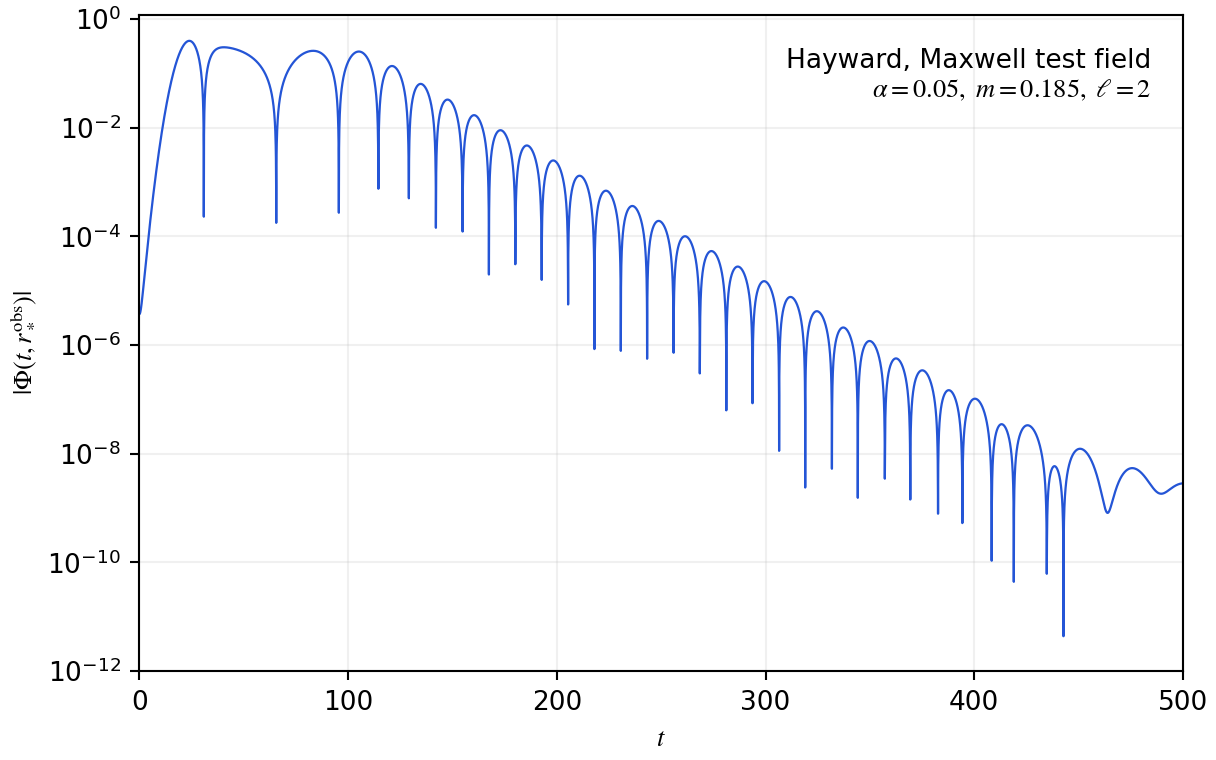} \qquad
\includegraphics[width=0.45\textwidth]{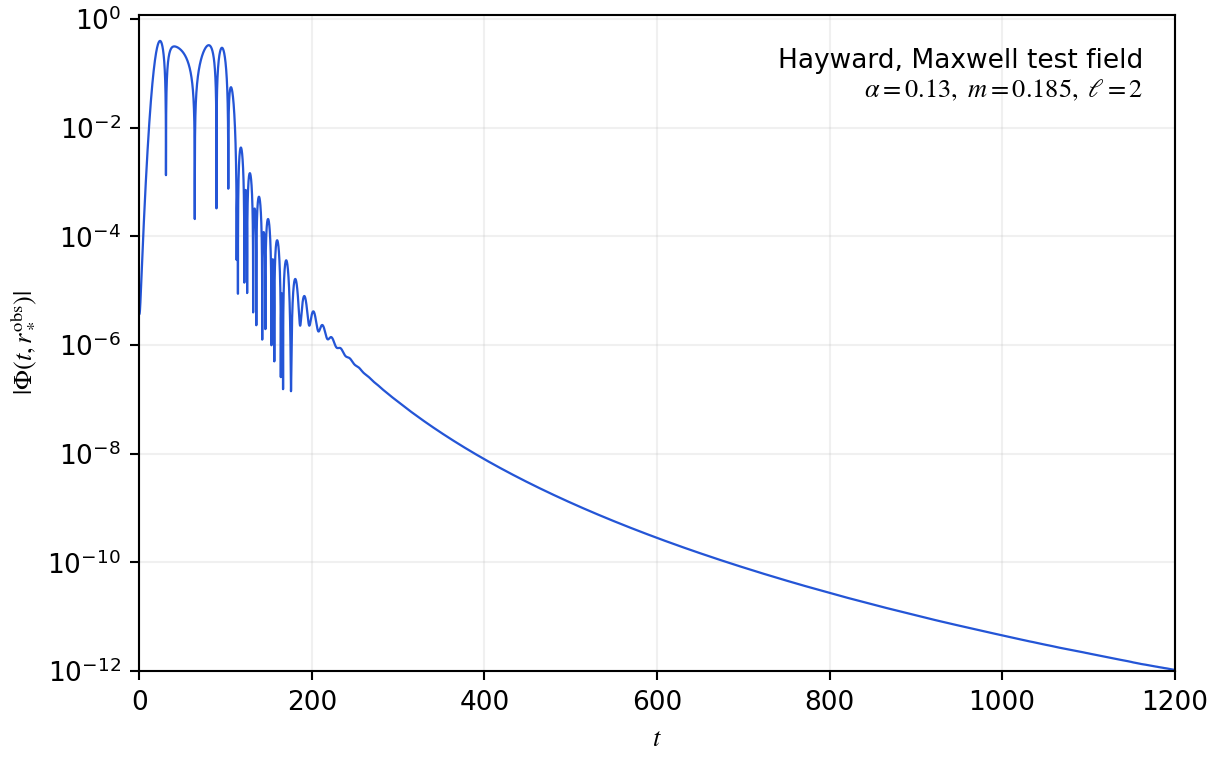} \qquad
\caption{ \it Absolute time-domain amplitude $|\Phi(t,r_*^{\rm obs})|$ on a logarithmic scale for the $\ell=2$ auxiliary Maxwell perturbation of Hayward black holes. The left and right columns show type-I and type-II solutions, respectively.}
\label{Haphi}
\end{figure}

Figures~\ref{Hsv} and \ref{Hav} show the scalar and auxiliary Maxwell potentials, respectively, and Figs.~\ref{Hsphi} and \ref{Haphi} show the corresponding time-domain profiles. As in the Bardeen family, the type-II solutions can develop a double-barrier potential and late-time echoes.

\begin{figure}[!htbp]
\centering
\includegraphics[width=0.45\textwidth]{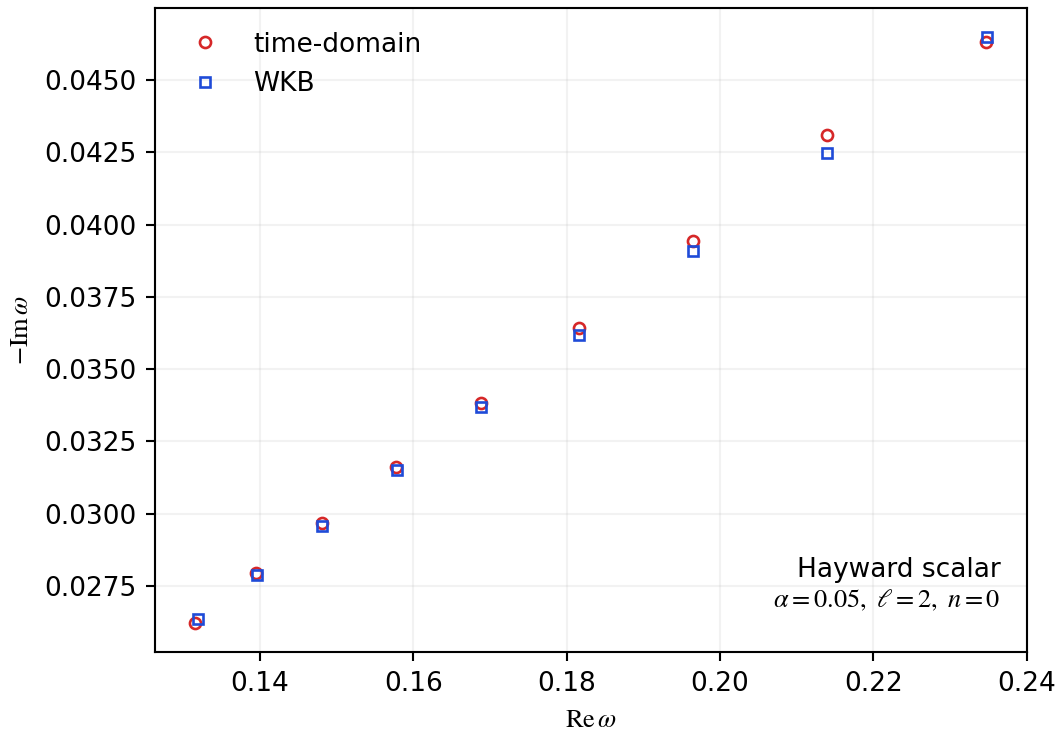} \qquad
\includegraphics[width=0.45\textwidth]{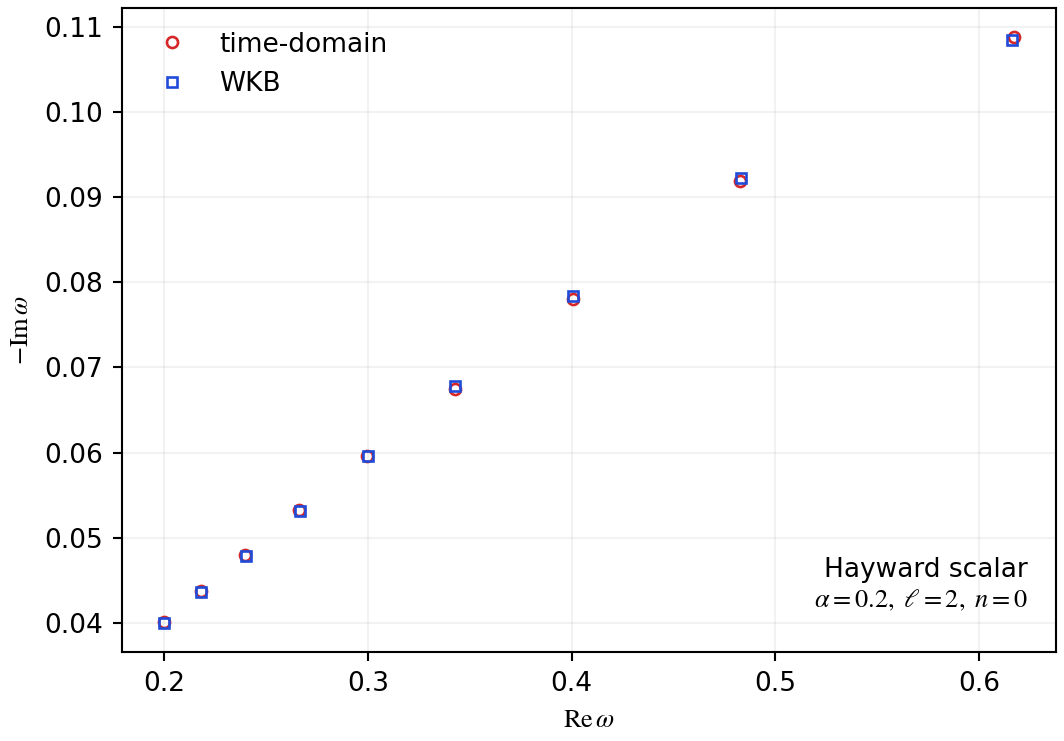} \qquad

\caption{ \it Positive damping rate $-\operatorname{Im}\omega$ versus oscillation frequency $\operatorname{Re}\omega$ for the fundamental scalar modes of Hayward black holes. Red circles and blue squares denote the time-domain and WKB results, respectively.}
\label{hms}
\end{figure}

\begin{figure}[!htbp]
\centering
\includegraphics[width=0.45\textwidth]{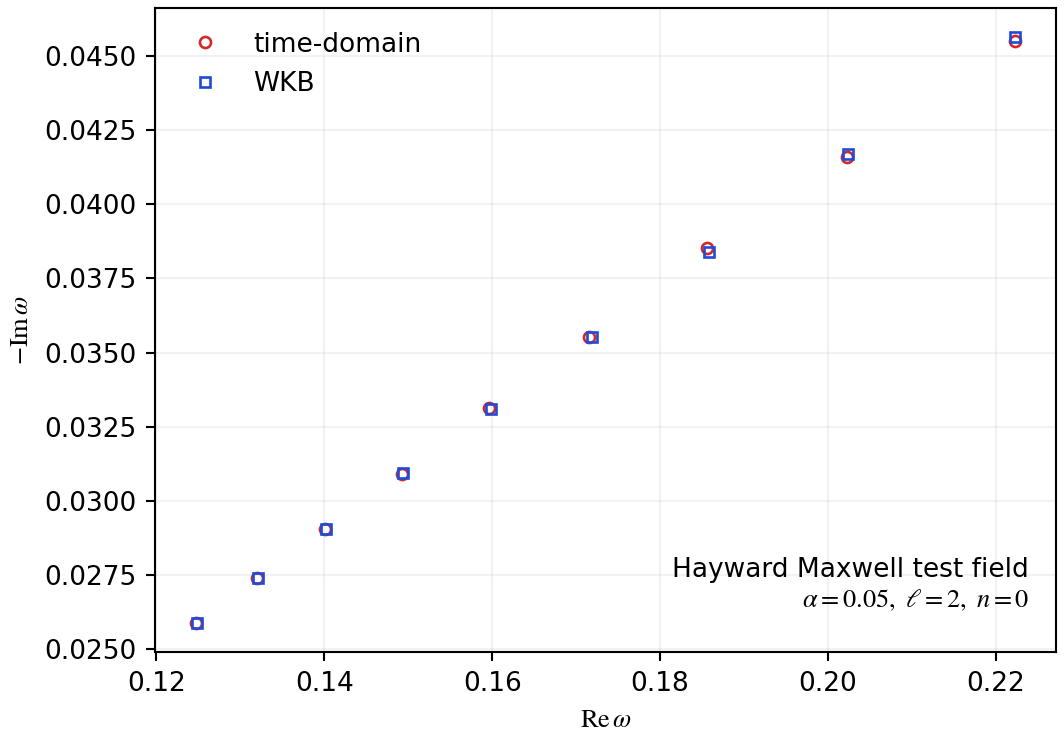} \qquad
\includegraphics[width=0.45\textwidth]{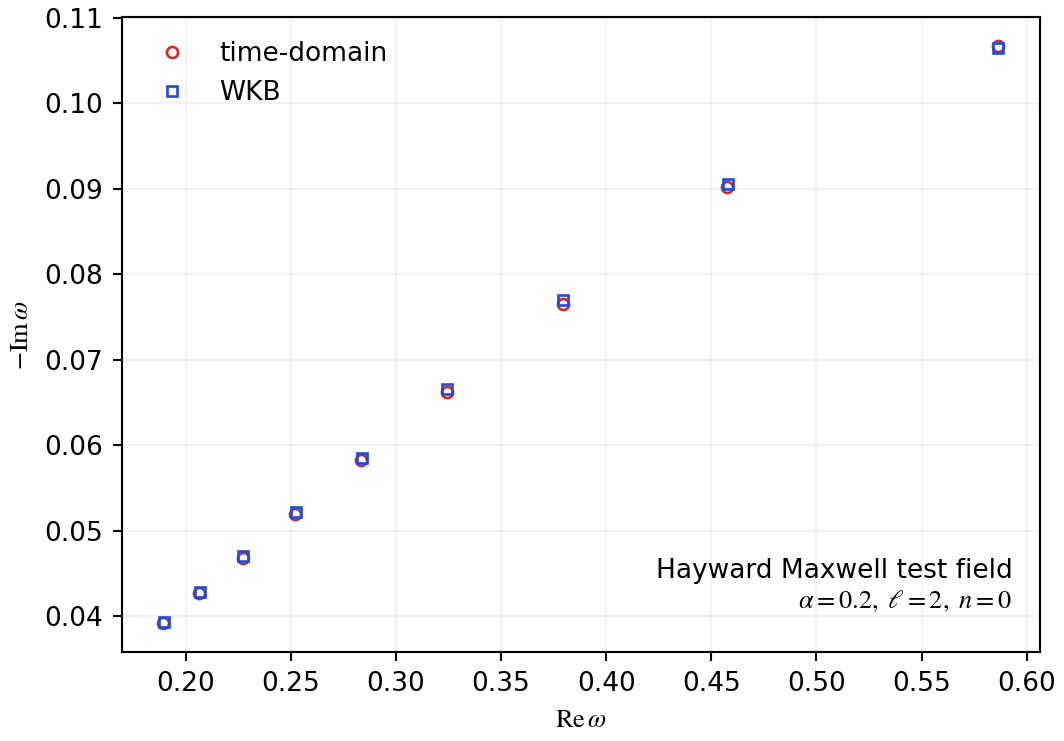} \qquad

\caption{ \it Positive damping rate $-\operatorname{Im}\omega$ versus oscillation frequency $\operatorname{Re}\omega$ for the fundamental auxiliary Maxwell modes of Hayward black holes. Red circles and blue squares denote the time-domain and WKB results, respectively.}
\label{hma}
\end{figure}

\FloatBarrier
We compute the Hayward QNM frequencies by both the WKB and time-domain methods. The results are listed in Tables~\ref{hqnms} and \ref{hqnma} and displayed in Figs.~\ref{hms} and \ref{hma}. The two methods agree closely for the single-barrier configurations represented in these tables.

\begin{table}[!htbp] 
\centering 
\caption{\it Fundamental ($n=0$) QNM frequencies of the $\ell=2$ scalar perturbation for Hayward black holes.}
\begin{tabular}{ccccc} 
\toprule
 &$\alpha=0.05$  & &$\alpha=0.2$\\
$m$ & WKB   & Time domain & WKB&Time domain\\
\midrule 
0.4&0.234849-0.046483i&0.234716-0.046300i&0.616309-0.108449i&0.616914-0.108779i\\
0.6&0.213892-0.042479i&0.214009-0.043104i&0.483121-0.092249i&0.482850-0.091917i\\
0.8&0.196411-0.039092i&0.196451-0.039448i&0.400458-0.078374i&0.400464-0.078016i\\
1.0&0.181596-0.036194i&0.181580-0.036411i&0.342757-0.067767i&0.342627-0.067512i\\
1.2&0.168872-0.033691i&0.168823-0.033836i&0.299861-0.059577i&0.299510-0.059545i\\
1.4&0.157823-0.031508i&0.157759-0.031615i&0.266616-0.053110i&0.266319-0.053192i\\
1.6&0.148137-0.029589i&0.148069-0.029675i&0.240054-0.047891i&0.239728-0.047983i\\
1.8&0.139574-0.027888i&0.139509-0.027961i&0.218329-0.043596i&0.218113-0.043695i\\
2.0&0.131949-0.026372i&0.131499-0.026226i&0.200221-0.040004i&0.200059-0.040089i\\

\\
\bottomrule 
\end{tabular}
\label{hqnms}
\end{table}

\begin{table}[!htbp] 
\centering
\caption{\it Fundamental ($n=0$) QNM frequencies of the $\ell=2$ auxiliary Maxwell perturbation for Hayward black holes.}
\begin{tabular}{ccccc} 
\toprule
 &$\alpha=0.05$  & &$\alpha=0.2$\\
$m$ & WKB   & Time domain & WKB&Time domain\\
\midrule	
0.4&0.222304-0.045639i&0.222277-0.045494i&0.586158-0.106439i&0.586446-0.106679i\\
0.6&0.202436-0.041708i&0.202325-0.041593i&0.458007-0.090577i&0.457609-0.090205i\\
0.8&0.185875-0.038382i&0.185607-0.038529i&0.379249-0.076953i&0.379261-0.076534i\\
1.0&0.171843-0.035538i&0.171559-0.035541i&0.324463-0.066538i&0.324483-0.066236i\\
1.2&0.159797-0.033080i&0.159601-0.033120i&0.283797-0.058496i&0.283553-0.058289i\\
1.4&0.149337-0.030937i&0.149223-0.030903i&0.252304-0.052147i&0.252157-0.051961i\\
1.6&0.140169-0.029052i&0.140036-0.029032i&0.227153-0.047022i&0.227069-0.046868i\\
1.8&0.132064-0.027383i&0.131945-0.027386i&0.206587-0.042805i&0.206518-0.042671i\\
2.0&0.124848-0.025894i&0.124726-0.025885i&0.189449-0.039278i&0.189367-0.039193i\\

\\
\bottomrule 
\end{tabular}
\label{hqnma}
\end{table}

\FloatBarrier

\section*{Appendix B: Null geodesics of the Hayward class}\label{app2}

This appendix presents the null-geodesic and light-ring results for the Hayward family.

\begin{figure}[!htbp]
\centering
\includegraphics[width=0.45\textwidth]{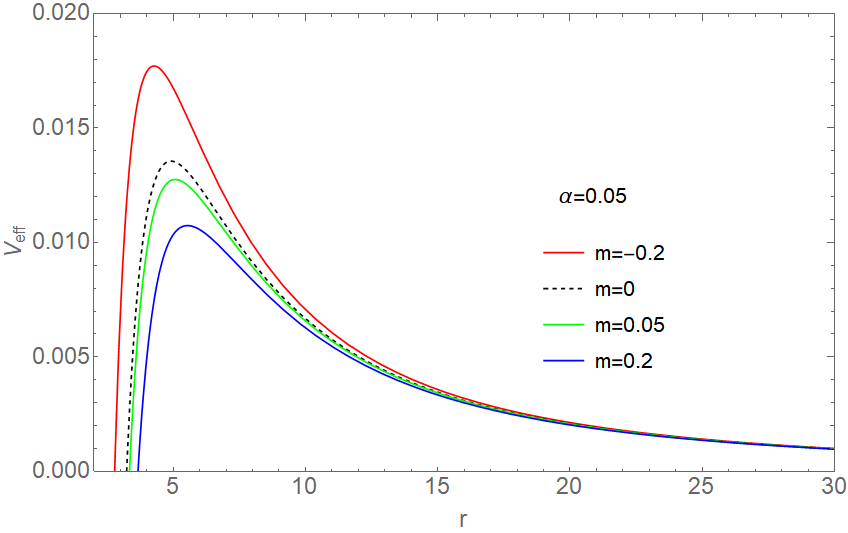} \qquad
\includegraphics[width=0.45\textwidth]{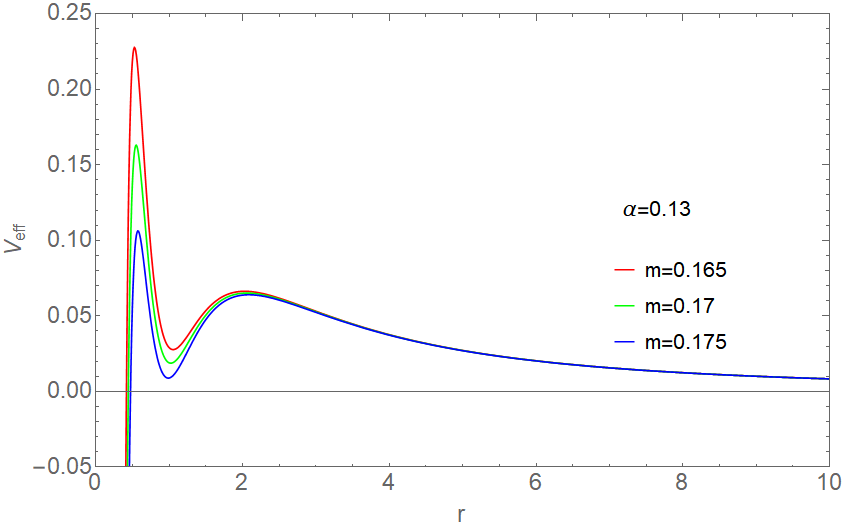}

\caption{ \it Null effective potential $V_{eff}$ for Hayward black holes. The left and right panels show type-I and type-II solutions, respectively.}
\label{hVeffnull}
\end{figure}

\begin{figure}[!htbp]
\centering
\includegraphics[width=0.45\textwidth]{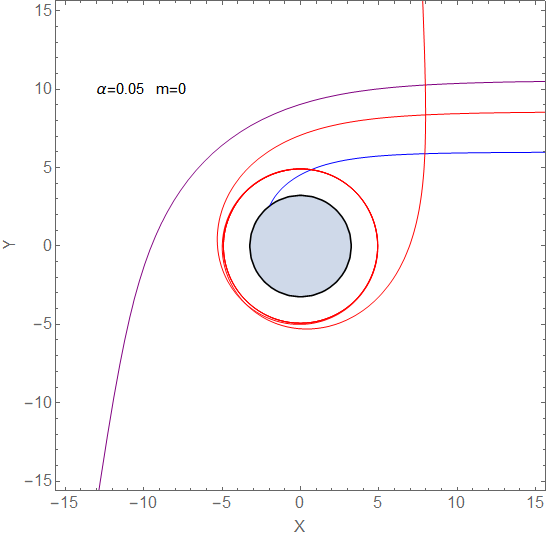} \qquad
\includegraphics[width=0.44\textwidth]{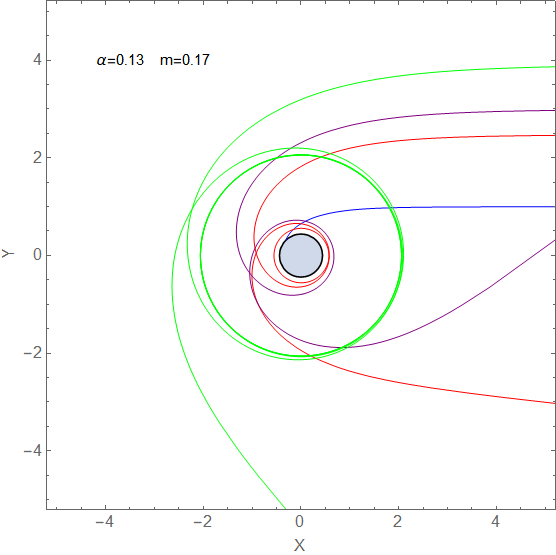}

\caption{ \it Representative null geodesics of Hayward black holes. The left panel shows a type-I black hole with one unstable light ring. The right panel shows a type-II black hole with two unstable light rings and an intervening stable circular null orbit.}
\label{hgeop}
\end{figure}

\begin{figure}[!htbp]
\centering
\includegraphics[width=0.45\textwidth]{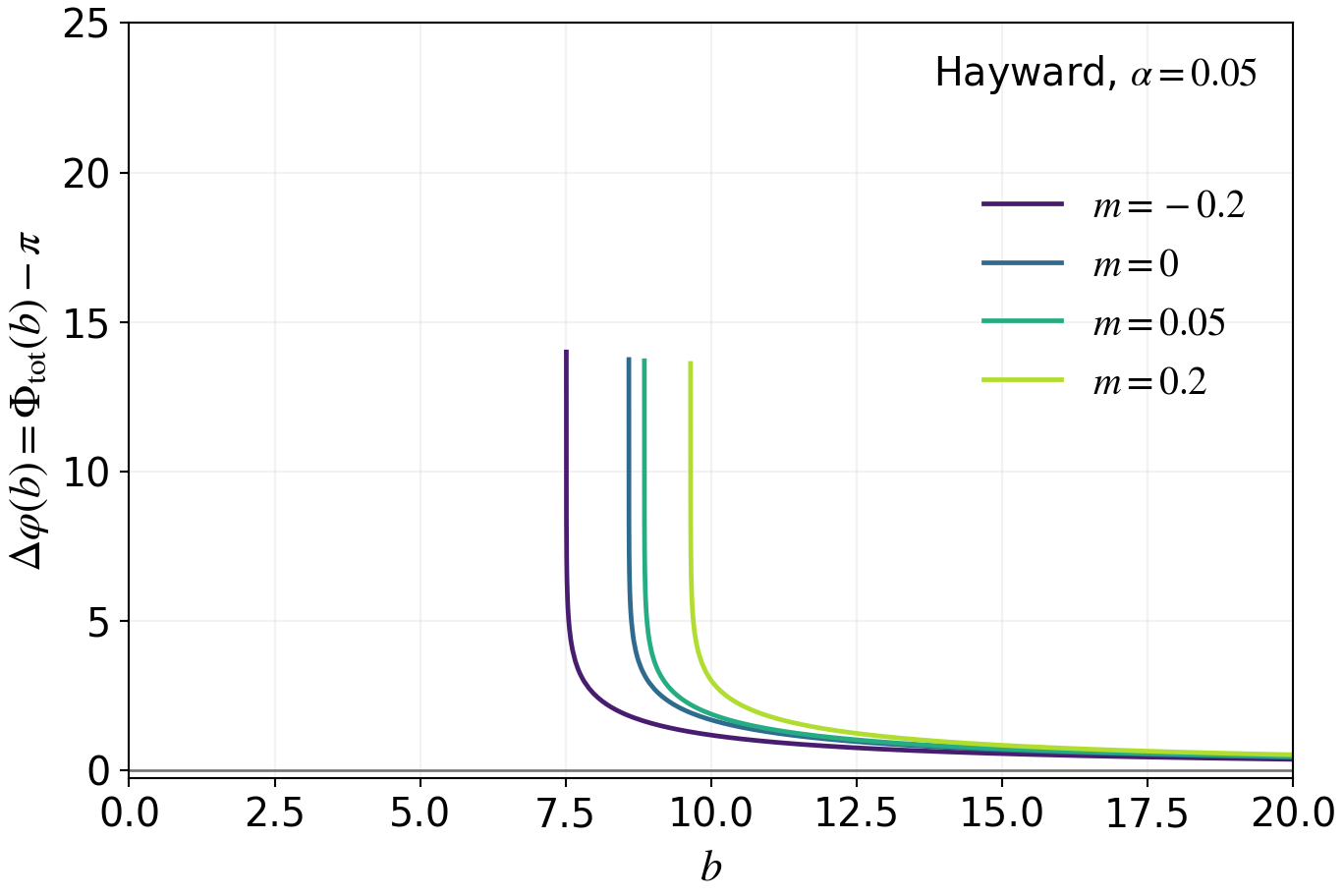} \qquad
\includegraphics[width=0.45\textwidth]{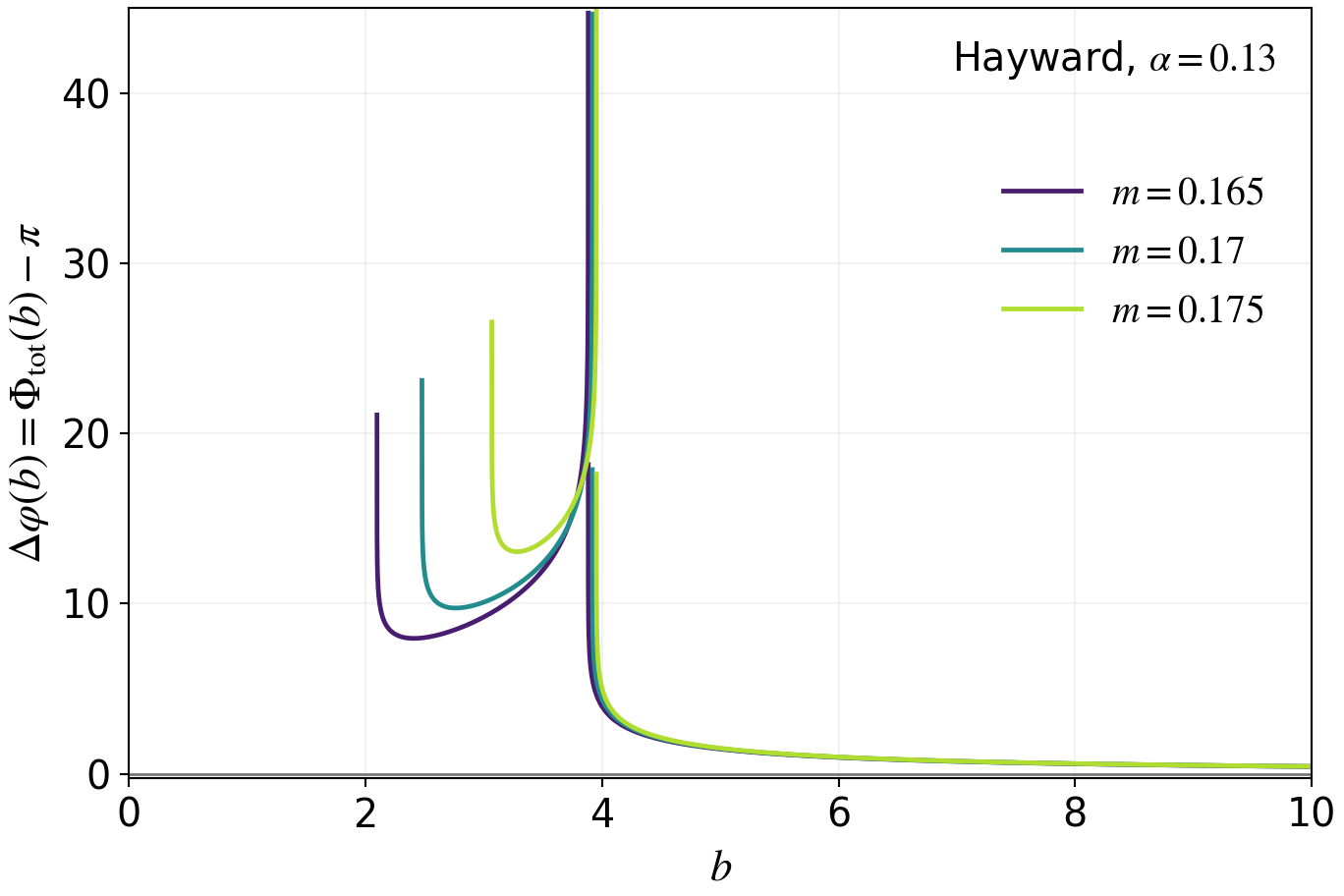}

\caption{ \it Deflection angle of returning null rays for Hayward black holes with $\alpha=0.05$ (left) and $0.13$ (right). Separate curve segments are the connected scattering branches delimited by accessible critical impact parameters; blank low-$b$ intervals denote capture.}
\label{htot}
\end{figure}

Figure~\ref{hVeffnull} shows the null effective potentials, and Fig.~\ref{hgeop} shows representative null geodesics of the background metric. The deflection angle $\Delta\varphi$ of returning rays is plotted as a function of $b$ in Fig.~\ref{htot}. The optical-metric qualification discussed in Sec.~\ref{4} applies here as well.

\section*{Acknowledgement}

We thank Mengyun Lai and Libo Xie for useful discussions. This work was supported by the National Natural Science Foundation of China (NSFC) under Grant Nos. 12205123 and 12565010 and by the Jiangxi Provincial Natural Science Foundation under Grant No. 20262BAC240348 and 20232BAB211029.

\FloatBarrier

\end{document}